\documentstyle[12pt,aaspp4]{article}
\begin{document}
\parindent=1.0cm

\title{A Near-Infrared Photometric Survey of Metal-Poor Inner Spheroid 
Globular Clusters and Nearby Bulge Fields} 

\author{T. J. Davidge \altaffilmark{1}}

\affil{Canadian Gemini Office, Herzberg Institute of Astrophysics,
\\National Research Council of Canada, 5071 W. Saanich Road,\\Victoria, 
B. C. Canada V8X 4M6\\ {\it email:tim.davidge@hia.nrc.ca}}

\altaffiltext{1}{Visiting Astronomer, Cerro Tololo Inter-American 
Observatory, which is operated by AURA, Inc., under contract from the 
National Science Foundation}

\begin{abstract}

	Images recorded through $J, H, K, 2.2\mu$m continuum, and CO filters 
have been obtained of a sample of metal-poor ([Fe/H] $\leq -1.3$) globular 
clusters in the inner spheroid of the Galaxy. The shape and color of the upper 
giant branch on the $(K, J-K)$ color-magnitude diagram (CMD), combined with the 
$K$ brightness of the giant branch tip, are used to estimate the metallicity, 
reddening, and distance of each cluster. CO indices are used to identify bulge 
stars, which will bias metallicity and distance estimates if not culled from 
the data. The distances and reddenings derived from these data are consistent 
with published values, although there are exceptions. The reddening-corrected 
distance modulus of the Galactic Center, based on the Carney et al. (1992, ApJ, 
386, 663) HB brightness calibration, is estimated to be $14.9 \pm 0.1$. 
The mean upper giant branch CO index shows cluster-to-cluster 
scatter that (1) is larger than expected from the uncertainties 
in the photometric calibration, and (2) is consistent with a 
dispersion in CNO abundances comparable to that measured among halo stars. The 
luminosity functions (LFs) of upper giant branch stars in the program clusters 
tend to be steeper than those in the halo clusters NGC 288, NGC 362, and NGC 
7089. The majority of inner spheroid clusters fall along the integrated $J-K$ 
versus metallicity relation defined by halo clusters; however, many of the  
inner spheroid clusters do not follow the relation between integrated CO index 
and metallicity measured for halo clusters, in that they have CO indices 
that are too small.

	Bulge fields were also observed near most clusters. The 
slope of the giant branch LF does not vary significantly between most fields, 
although the LFs in Baade's Window and near NGC 6273 are significantly 
shallower than average. Metallicities estimated from the slope of the 
upper giant branch on the $(K, J-K)$ CMDs of fields within 6 degrees 
of the Galactic Center are consistent with previous studies.
Finally, the data suggest that the HB content may not be uniform 
thoughout the bulge, in the sense that a larger than average number of 
red HB stars may occur in fields closest to the Galactic Center.

\end{abstract}

\keywords{globular clusters: general -- Galaxy: structure -- Galaxy: center -- 
infrared: stars -- stars: late type -- stars: horizontal branch}

\section{INTRODUCTION}

	While the inner spheroid of the Galaxy is dominated by stars with 
metallicities near solar (e.g. McWilliam \& Rich 1994, Geisler \& Friel 
1992, Ratag et al. 1992), there is also a modest 
population of stars with [Fe/H] $\leq -1$. These metal-poor 
stars are potentially of great cosmological importance since, if the first 
episodes of star formation occured near the center of the present-day Galaxy, 
where the density of gaseous material, and hence the incidence of 
cloud-cloud collisions, may have been greatest during early epochs, then they 
should be among the oldest objects in the Galaxy (e.g. Larson 1990). 

	Deep photometric surveys of globular clusters at low Galactic latitudes 
will ultimately provide the most reliable means of charting the early 
evolution of the inner spheroid. A critical first step in the 
study of these objects is to determine basic cluster properties, such as 
metallicity, reddening, and distance so that targets can be selected for deep 
imaging surveys. While many low latitude clusters already have reddening, 
metallicity, and distance estimates, in the majority of cases these quantities 
were measured from data recorded at visible wavelengths, and hence are 
subject to uncertainties introduced by (1) the large and spatially 
variable levels of extinction prevalant at low Galactic latitudes, and 
(2) contamination from bulge stars, which elevates crowding levels and can 
frustrate efforts to isolate cluster star samples. In addition, some 
low-latitude clusters lack CMDs, adding further uncertainty to 
metallicity, reddening, and distance estimates.

	The complications introduced by reddening 
can be reduced by observing at infrared wavelengths, as 
the extinction in $K$ is roughly one-tenth that in $V$ 
(Rieke \& Lebofsky 1985). The effects of field star contamination are also 
reduced in the infrared, since the contrast between the brightest 
red giants and the unresolved, relatively blue, body of the cluster is 
enhanced with respect to visible wavelengths, so that the richly populated 
central regions of clusters, where field star contamination is minimized, can 
be surveyed for bright stars from ground-based facilities using conventional 
observing techniques. 

	In the present study broad and narrow-band near-infrared observations 
are used to estimate the reddenings, metallicities, and distances of all 
globular clusters that, according to the 
1996 version of the Harris (1996) database, have 
[Fe/H] $\leq -1.3$ and R$_{GC} \leq 3.5$ kpc. Metallicities and reddenings are 
determined from the shape and color of the upper giant 
branch on $(K, J-K)$ CMDs, while distances are determined from the $K$ 
brightness of the red giant branch (RGB) tip. Two important aspects 
of this survey are that (1) the data were recorded with a single instrumental 
setup during a 5 night observing run, thereby creating a homogeneous 
database that is free of the errors that 
might occur when observations from different runs and/or 
instruments are combined, and (2) narrow-band 
CO indices are used to identify bright bulge stars, which will influence 
the measured cluster properties if not removed from the data.

	The clusters meeting the selection criteria are listed in Table 1. 
Djorgovski 1 and HP 1 were originally observed as part of a 
preliminary metal-rich cluster survey that was also conducted during this 
observing run. However, subsequent investigation 
indicated that these clusters are metal-poor (Davidge 2000), and so they were 
added to the current sample. The observations of NGC 6139 and NGC 6287 
were discussed previously by Davidge (1998).

	The sample listed in Table 1 is biased against heavily obscured 
clusters, which will predominate at the lowest Galactic latitudes. Evidence 
for such a bias comes from the projected distribution of 
the target clusters, which is shown in Figure 1. The clusters do not 
follow the contours of the bulge, which has an axial ratio $\sim 0.7$ 
(Blanco \& Terndrup 1989); rather, the clusters tend to lie along the minor 
axis of the bulge and there is a dearth of clusters along the major axis. If 
it is assumed that there is an isotropically distributed metal-poor cluster 
population in the innermost regions of the Galaxy, then a number of 
clusters evidently await discovery, especially at Galactic latitudes $\leq 
10^o$ (e.g. Frenk \& White 1982).

	The observations and the procedures used to reduce the data are 
described in \S 2. The photometric measurements, luminosity functions (LFs), 
color-magnitude diagrams (CMDs), RGB-tip brightnesses, and integrated color 
measurements are discussed in \S 3, as is the identification of metal-rich 
bulge stars. The distances, reddenings, metallicities, and near-infrared 
spectral energy distributions (SEDs) of the target clusters are investigated 
in \S 4. The LFs of upper giant branch stars in bulge fields, which were 
observed to sample the background stellar component near the target clusters, 
are examined in \S 5. A summary and discussion of the results follows in \S 6.

\section{OBSERVATIONS AND REDUCTIONS}

	The data were obtained with the CIRIM camera, which was mounted at the 
Cassegrain focus of the CTIO 1.5 metre telescope. CIRIM contains a $256 \times 
256$ Hg:Cd:Te array, with an angular scale of 0.6 arcsec per pixel at this 
focus, so that each exposure covers a $154 \times 154$ arcsec field.

	Two fields were typically observed in and around each cluster, and 
these sampled the cluster center (Field 1), and an area 2 arcmin North (Field 
2). A third field, offset 30 arcmin North of each cluster (Field 3), was also 
observed to sample the properties of bulge stars near each cluster, and conduct 
a serendipitous survey of the bulge stellar content. At extremely low Galactic 
latitudes, where the density of bulge stars may change rapidly with distance 
from the Galactic Center (GC), and field-to-field differences in reddening 
can be significant, the Field 3 data provide only a 
crude measure of the background in the vicinity of each cluster. Only 
Field 1 was observed for Djorgovski 1, while Field 3 was not observed for NGC 
6717. The central regions of the halo clusters NGC 288, NGC 362, and NGC
7089 were also observed.

	$J, H,$ and $K$ images were obtained of all fields, while CO and 
$2.2\mu$m continuum images were also recorded of each Field 1 so that the 
strength of $2.3\mu$m CO absorption could be measured in moderately bright ($K 
\leq 13$) stars. A complete observing sequence consisted of four exposures per 
filter, with the telescope offset between each of these to create a 
$5 \times 5$ arcsec square dither pattern. Integration times for 
Field 1 were kept to between 5 and 15 seconds per dither position for the 
broad-band filters to prevent saturating the brightest stars. 
The exposure times for Fields 2 and 3 were 60 sec per dither position. 
The image quality, measured from the processed images, was typically 
1.2 to 1.5 arcsec FWHM independent of wavelength, although 
image quality better than 1.2 arcsec would not be detected as such due to the 
0.6 arcsec pixel sampling.

	The data were reduced using the procedures described by Davidge \& 
Courteau (1999a). The processing sequence consisted of: (1) 
linearization, using coefficients supplied by CTIO staff, (2) dark subtraction, 
(3) flat-fielding, using dome flats, (4) removal of thermal emission 
signatures and interference fringes, using calibration frames constructed 
from background sky fields, and (5) subtraction of the 
DC sky level, which was estimated by calculating the mode of the pixel 
intensity distribution for each image. The processed images for each field 
were then aligned to correct for the offsets introduced during acquisition, and 
the results were median-combined on a filter-by-filter basis to reject cosmic 
rays and bad pixels.

\section{PHOTOMETRIC PROPERTIES}

\subsection{Photometric Measurements}

	The photometric calibration is based on observations of standard stars 
from Elias et al. (1982) and Casali \& Hawarden (1992), which 
have $J-K$ ranging between --0.2 and 0.9. A total of 31 standard star 
observations were obtained. Casali \& Hawarden (1992) give standard 
brightnesses in the UKIRT system, and these were transformed into the CIT/CTIO 
system using the relations listed by these authors. The estimated uncertainty 
in the photometric zeropoints, based on the scatter in the standard star 
measurements, is $\pm 0.02$ mag in $J, H, K$, and the CO index. 
The scatter in the calibration is demonstrated in Figure 2, which shows 
the difference between the standard and instrumental brightness 
in $K$ as a function of instrumental $J-K$ color. The scatter in this figure 
is comparable to what has been achieved in other near-infrared photometric 
studies with array detectors (e.g. Davidge \& Harris 1995), and indicates that 
the error in a single observation is on the order of a few hundredths of a 
magnitude.

	Stellar brightnesses were measured with the PSF-fitting routine ALLSTAR 
(Stetson \& Harris 1988), using star lists and PSFs obtained from 
DAOPHOT (Stetson 1987) tasks. Aperture corrections were derived 
from the 30 -- 50 bright stars in each frame that were used to construct the 
PSFs, after removing all other detected stars from the image.

	The majority of the inner spheroid clusters are located in densely 
populated bulge fields, and contamination from field stars complicates efforts 
to measure integrated colors and trace the cluster giant branch on the CMD. To 
extract a dataset for each cluster in which the ratio of cluster to bulge stars 
is relatively large, the radius at which the surface brightness profile of the 
underlying body of each cluster equals that of the surrounding 
background was estimated from the Field 1 $K$ images. These measurements were 
made after subtracting all detected stars from the images and 
then smoothing the result with a $6 \times 6$ arcsec top hat median filter to 
suppress undetected stars and artifacts of the subtraction process. For the 
remainder of the paper the regions interior to and exterior to this 
radius will be referred to as the `inner cluster' and `outer cluster' 
fields, respectively. For roughly one-third of the clusters 
the surface brightness profile exceeded that of the background over all 
of Field 1, and in these cases the `inner cluster' designation was assigned 
to Field 1 in its entirety, and an `outer cluster' region was not defined. The 
inner cluster radii are listed in the last column of Table 1; those 
clusters without an outer cluster field have a dash entered in this column.

\subsection{Luminosity Functions}

	The LFs of NGC 288, NGC 362, and NGC 7089 are shown in Figure 3, and 
these follow power-laws at the bright end. The HB produces a bump in the NGC 
362 LF near $K = 13.5$, while the onset of the subgiant branch (SGB) is evident 
near the faint end of the NGC 288 and NGC 362 LFs; the main sequence turn-off 
occurs near $K \sim 17.5$ in NGC 288 (Davidge \& Harris 1997). 
A crude measure of the completeness limit for each dataset 
can be estimated from the point at the faint end where the number counts start 
to decline, and the data in Figure 3 are thus complete when $K \geq 16 - 17$.

	To quantify the slopes of the LFs, the method of least squares was 
used to fit power-laws in the brightness interval between the 
RGB-tip and the point at which incompleteness sets in (NGC 7089) or 
the onset of the SGB (NGC 288 and NGC 362); the HB was also omitted from the 
NGC 362 data. The exponents, $x$, derived for these clusters 
are compared in Table 2, and these agree at the $2\sigma$ level. The 
mean value of $x$, with each value weighted according to the reciprocal of 
the estimated uncertainty, is 0.17. 

	The $K$ LFs of the inner cluster fields, which are shown in Figure 
4, also follow power-laws at the bright end. The Field 3 $K$ LFs, 
which are truncated by detector saturation at $K = 10$, are shown as dashed 
lines in Figure 4, and it is evident that bulge stars contribute 
significantly to the number counts in many of the inner cluster fields. 
The Field 3 LFs, which are discussed in more detail in \S 5, 
also follow power-laws, and in many cases appear to be 
steeper than the cluster LFs. 

	The degree of contamination from bulge stars varies substantially from 
cluster-to-cluster. To quantify this contamination, the number densities of 
stars with $K$ between 10 and 12.5 in the inner cluster and bulge fields were 
counted and then ratioed to form a contamination index $C$, and the results are 
listed in the last column of Table 3. The brightness 
interval for computing $C$ was chosen to bracket the saturation limit of the 
Field 3 data ($K = 10$) and the approximate completeness limit for clusters 
in very crowded fields ($K = 12.5$), such as HP 1. The entries in Table 3 
indicate that bulge stars contribute at least 10\% (i.e. $C \leq 10$) 
of the bright stellar content in the majority of inner cluster fields.

	The method of least squares was used to fit power-laws to each of the 
inner cluster fields after subtracting the Field 3 LF, and the 
exponents computed in this manner are listed in the second column of Table 3. 
The fits were restricted to the brightness interval between $K = 10$ and 12.5. 
Despite being restricted to the upper end of each giant branch, where the 
number of stars is modest, in the majority of cases the best fit power-law 
tracks the fainter portions of the giant branch. The weighted mean exponent 
is $\overline{x} = 0.26 \pm 0.01$, indicating that the giant branches of 
the inner spheroid clusters are, on average, significantly 
steeper than those in the halo clusters. 
The exponent for NGC 6287 is significantly steeper than the mean, although 
the LF of this cluster becomes noticeably flatter when $K \geq 12.5$.

\subsection{$(K, J-K)$ CMDs}

	The $(K, J-K)$ CMDs of NGC 288, NGC 362, and NGC 7089, based on 
stars that were detected in all 5 filters, are plotted in Figure 5. The 
cluster giant branches are clearly evident, as is the HB of NGC 362 near 
$K = 13.5$. The solid lines show the locus of the NGC 362 observations 
published by Frogel, Persson, \& Cohen (1983), and the normal points for 
NGC 288 listed in Table 2 of Davidge \& Harris (1997). The current data are 
in good agreement with these published observations. 
The brightest star detected near the center of NGC 288 is $\sim 1.5$ fainter 
than the RGB-tip of this cluster, based on the sample of stars observed 
by Frogel, Persson, \& Cohen (1983). The present data do not 
sample bright upper giant branch stars in NGC 288 due to the 
relatively low central stellar density of this cluster.

	The CMDs of NGC 288 and NGC 7089 are dominated by first ascent giants. 
However, a prominent asymptotic giant branch (AGB) sequence, which dominates 
when $K \leq 9.6$, is seen $\sim 0.05$ mag blueward of the RGB in the NGC 362 
CMD. The AGB-tip brightness infered from the NGC 362 data is $\sim 0.3$ mag 
brighter in $K$ than the brightest star observed by Frogel, Persson, \& Cohen 
(1983). A prominent AGB sequence can also be seen above the HB in the $(V, 
B-V)$ CMD of NGC 362 obtained by Harris (1982), which is based on observations 
of stars in the outer regions of the cluster.

	The $(K, J-K)$ CMDs of the inner cluster and bulge fields are 
shown in Figure 6. Only stars detected in all five filters are plotted 
in the cluster CMDs, while the bulge field CMDs contain stars 
detected in all three broad-band filters. The quality of the cluster CMDs is 
sensitive to field star contamination: the clusters with $C \geq 10$ tend to 
have tight, well-defined CMDs, while the scatter increases when $C \leq 10$. 
The CMDs of Djorgovski 1 ($C = 2$), HP 1 ($C = 4$), and NGC 6522 
($C = 3$) show two sequences, one belonging to the cluster, the other to the 
bulge. Differential reddening likely also contributes to the scatter in 
the CMDs of some clusters, such as NGC 6287 and NGC 6626.

	The bulge field CMDs are smeared by a combination of depth effects and 
the broad range of metallicities among bulge stars. The red HB clump of the 
bulge can be seen in the CMDs of some bulge fields (e.g. HP 1, NGC 6453, and 
NGC 6558), and in fields with only moderate amounts of crowding there is the 
impression that the main sequence turn-off of the bulge has been detected. 
Blue stars are also seen in some of the bulge CMDs, and these 
are a combination of bulge HB and foreground disk objects.

	Minniti, Olszewski, \& Rieke (1995a) obtained $(K, J-K)$ CMDs of 
nine of the clusters studied here, while deep $J$ and $K$ 
observations of NGC 6626 were discussed by Davidge, C\^{o}t\'{e}, 
\& Harris (1996), and the trends defined in these studies are compared with 
the current CMDs in Figure 6. With the exception of NGC 6293 and NGC 
6333, there is a tendency for the Minniti et al. (1995a) data to have $J-K$ 
colors that are 0.2 to 0.4 mag larger than the current measurements.

	The giant branch sequences shown by Minniti et al. (1995a) are not 
consistent with other published observations. For example, the M92 observations 
made by Minniti et al. (1995a) have $J-K$ colors that are substantially larger 
than those measured by Cohen, Frogel, \& Persson (1978). This is most evident 
near the RGB-tip, which Minniti et al. place near $J-K \sim 1.2$, compared with 
$J-K \sim 0.7$ measured by Cohen et al. (1978) and Davidge \& Courteau (1999a). 
The Minniti et al. (1995a) CMD of M22 also has $J-K$ colors that are 0.1 -- 0.2 
mag larger than the giant branch sequences defined by Davidge \& Harris (1996) 
and Frogel, Persson, \& Cohen (1983). Uncertainties in the Minniti et al. 
(1995a) color calibration would explain the substantial scatter in their 
Figure 4.

\subsection{CO Indices and the Identification of Bright Bulge Stars}

	The effective wavelengths of the CO and $2.2\mu$m continuum filters 
differ by less than $0.1\mu$m, and so the CO index is very insensitive to 
reddening variations, with $\frac{E(CO)}{E(J-K)} = -0.08$ (Elias, Frogel, \& 
Humphreys 1985; Rieke \& Lebofsky 1985). This modest reddening dependence, 
combined with the metallicity sensitive nature of the $2.3\mu$m CO bands, 
makes the CO index a powerful tool for identifying metal-rich bulge stars, 
which can be difficult to identify solely from broad-band colors. 

	The $(K, CO)$ CMDs for stars in NGC 288, NGC 362, and NGC 7089 are 
shown in Figure 7. The upper giant branches of these clusters define 
almost vertical sequences on the $(K, CO)$ CMD, even near the RGB-tip. The CO 
indices of upper giant branch stars in these clusters also show a marked 
metallicity dependence: stars in NGC 7089, which is the most metal-poor 
cluster of the three, tend to have the smallest CO indices, while stars 
in NGC 362, which is the most metal-rich, tend to have the largest CO values. 
Stars in NGC 288 have CO indices that fall almost midway between the other 
two clusters.

	The $(K, CO)$ CMDs of the inner cluster fields are shown in the top 
panels of Figure 8, while the histogram distributions of CO indices for stars 
with $K \geq 12.5$, where the observational scatter in the CO indices 
is modest, are shown in the lower panels of Figure 8. The clusters with the 
lowest $C$ values tend to have the broadest CO distributions, due to 
contamination from bulge stars, and in some cases the CO distribution is 
bimodal. The clusters with the highest $C$ values, where bulge star 
contamination is smallest, have the tightest CO distributions.

	To identify metal-rich bulge stars a critical CO index was found for 
each cluster such that stars with CO $\geq$ CO$_{Crit}$ are bulge objects 
while those with CO $\leq$ CO$_{Crit}$ are likely cluster members. 
CO$_{Crit}$ was determined in two different ways, based on whether or not an 
outer cluster field was identified. Clusters having both inner and outer 
cluster fields, which tend to be those with the smallest $C$ values, are 
considered first. 

	The outer cluster fields are, by definition, dominated by 
bulge stars, so the CO distribution of these fields 
should approximate that of the background. The CO distributions of the inner 
and outer cluster fields, scaled according to field size, 
were subtracted and the CO index for which an excess number of stars in the 
inner cluster field occured with respect to the outer cluster field was 
determined. The stars with CO indices greater than this value in both the 
inner and outer cluster fields were rejected as bulge objects, and the 
remaining stars in both fields were assumed to be cluster members. 

	In practice only a modest number of stars were rejected using this 
criterion, although significant rejection rates 
occured for Djorgovski 1, HP 1, NGC 6522, and NGC 6558. The CMDs of these 
clusters before and after the removal of bulge stars are compared in Figure 9, 
and it is evident that the removal of bulge stars greatly affects 
the CMDs of these clusters near the bright end. The rejection 
of stars with large CO indices noticeably reduces the scatter in the $(K, J-K)$ 
CMDs, and this is most evident in the NGC 6522 dataset. 

	The differencing technique described above can not be applied to 
clusters where an outer cluster field was not defined. However, the $(K, CO)$ 
CMDs of the clusters that lack an outer cluster field 
tend to be relatively well-defined with only modest amounts of 
scatter, so that stars that deviate significantly from the cluster 
locus can be readily identified. For these clusters, the envelope of CO indices 
on the $(K, CO)$ CMD was defined by eye, and stars falling to 
the right of this envelope were rejected as bulge stars. Only a few stars were 
rejected per cluster.

	The rejection techniques described above identify only 
those stars that are significantly more metal-rich than 
the cluster, so that metal-poor bulge stars will not be rejected. Lacking 
additional information, such as the space motions of stars in these fields, 
the removal of metal-poor bulge stars is problematic. However, the 
metallicity distribution function of the bulge peaks near solar values, with 
the majority of objects having [M/H] $> -1$ (e.g. McWilliam \& Rich 1994). 
Hence, while the techniques applied here do not identify 
metal-poor bulge stars, the density of these objects is expected to be small 
compared with that of cluster stars. 

\subsection{RGB-tip Brightnesses}

	The HB is an almost vertical sequence on the infrared CMDs of 
metal-poor clusters (e.g. Davidge \& Courteau 1999a), and so is not well-suited 
as either a reference point for the vertical registration of 
CMDs or as a distance indicator at these wavelengths. If a cluster contains 
RR Lyrae variables with known periods then the infrared period - luminosity 
relation (Carney, Storm, \& Jones 1992; Longmore 
et al. 1990) can be applied, although for low latitude 
clusters contamination from field RR Lyraes must then be considered.

	The RGB-tip provides an alternate distance indicator in the infrared 
that, at least for clusters that are significantly more metal-poor than the 
dominant bulge component, can be corrected for field star contamination. A 
potential concern is that evolution on the upper giant branch proceeds at a 
relatively rapid pace, so that stars near the RGB-tip are rare when compared 
with earlier phases of evolution, and this may introduce biases in the RGB-tip 
brightness measurement. However, numerical simulations predict that modest 
surveys of bright stellar content are sufficient to estimate the 
RGB-tip brightness in globular clusters to $\pm 0.1$ mag 
(Crocker \& Rood 1984), and this prediction has been verified observationally 
(e.g. Figure 6 of Frogel, Cohen, \& Persson 1983).

	The $K$ brightness of the RGB-tip was estimated for each cluster 
by identifying the brightest star on the giant branch locus after 
removing metal-rich bulge stars, and the results are listed in the 
second column of Table 4. The stellar density near the center of NGC 288 is 
relatively low, and the field studied here is devoid of stars on the upper 
portions of the RGB. Therefore, the RGB-tip brightness for NGC 288 in Table 4 
is based on the observations made by Frogel, Persson, \& Cohen (1983).

	The histogram distribution of RGB-tip brightnesses, shown 
in the top panel of Figure 10, is skewed slightly towards systems with larger 
values of K$_{RGBT}$, due to a small number of clusters with significantly 
higher than average reddenings. The mean RGB-tip brightness is 
$\overline{K_{RGBT}} = 9.4$, with a standard deviation of 0.6 mag. 

\subsection{Integrated Near-Infrared Colors}

	Integrated colors provide another means of investigating the stellar 
content and chemical composition of globular clusters, although contamination 
from bright field stars is an issue at low Galactic latitudes (Davidge 2000, 
Bica et al. 1998). For the current study, integrated color measurements were 
restricted to the inner cluster fields, with the sky background measured in 
the outer cluster field. For those clusters where an 
outer cluster field was not defined, the colors were measured in a 51 arcsec 
radius annulus centered on the cluster, with the background measured 
in a 20 arcsec wide annulus at the edge of each frame, thereby sampling an 
area comparable to that of the central aperture. This latter procedure was also 
applied to NGC 362 and NGC 7089. Integrated colors could not be measured for 
NGC 288 because of the diffuse nature of this cluster. The integrated colors 
of the clusters considered in this paper are listed in Table 4. 
The integrated colors for Djorgovski 1 and HP 1 are those 
listed in Table 4 of Davidge (2000).

\section{CLUSTER PROPERTIES}

\subsection{Metallicities and Reddenings}

	Metallicities and reddenings were estimated by 
comparing the upper portions of the giant branch on the $(K, J-K)$ CMD with the 
16 Gyr Bergbusch \& VandenBerg (1992) isochrones. Normal points were generated 
for each cluster by calculating the mean $J-K$ color in $\pm 0.25$ mag bins 
along the $K$ axes of the $(K, J-K)$ CMDs and applying a $2.5\sigma$ 
rejection criteria to remove outliers. The Bergbusch \& VandenBerg 
models were selected as reference sequences because (1) they 
were computed with up-to-date input physics, and include the 
effects of oxygen-enhancement (but not enhancement of other $\alpha$ elements), 
(2) they span a wide range of metallicities, with a sample interval 
$\Delta$[Fe/H] $\sim 0.2$ dex at the metal-poor end, and (3) near-infrared 
observations of cluster stars were used to fine-tune the isochrones near the 
RGB-tip (VandenBerg 1992).

	Bergbusch \& VandenBerg (1992) list $V$ and $B-V$ pairs at various 
points along each isochrone, and these were transformed into $K$ and $J-K$ 
using the relations between visible and infrared colors listed in Table 5 of 
Davidge \& Harris (1995), which were derived from the giant branches of 
globular clusters with [Fe/H] between $-1.5$ 
and $-2.0$. The transformed isochrones were 
shifted to the apparent distance of each cluster using the RGB-tip 
brightnesses listed in Table 4 and, with the vertical placement thus fixed, the 
sequences were translated along the horizontal axes to fit the cluster giant 
branch. The metallicity adopted for each cluster is that of the isochrone 
that best fits the normal point sequence, while the net translation along 
the color axis for the best fitting isochrone gives $E(J-K)$.

	The metallicities and reddenings predicted for NGC 288, NGC 362, and 
NGC 7089 are listed in the second, third, and fourth columns of Table 5, while 
the corresponding quantities from Harris (1996) are listed in the last two 
columns. The NGC 288 CIRIM observations do not sample the upper portion of 
the giant branch, which is the portion of the CMD that contains the 
most significant information for metallicity determinations, 
and so the data for this cluster were supplemented with normal points derived 
from the aperture measurements made by Frogel, Cohen, \& Persson (1983).

	The metallicities derived from the isochrones 
are systematically smaller than those listed by Harris (1996). To 
investigate if this trend is unique to the current observations, 
the metallicity of M13 was estimated using normal points 
derived from the measurements published by Cohen et al. (1978) and 
Davidge \& Harris (1995), and the result is shown in the last row of Table 5. 
The resulting metallicity estimate for M13 is smaller than that given by Harris 
(1996), by an amount that is consistent with the NGC 288, NGC 362, and NGC 7089 
values. The mean offset from the Harris (1996) metallicities for the four 
clusters listed in Table 5 is $0.4 \pm 0.1$, and all subsequent metallicity 
estimates derived from the shape of the upper giant branch will be adjusted by 
this amount.

	The systematic difference between the metallicity estimates in 
Table 5 occurs because the models overestimate the 
slope of the upper giant branch. VandenBerg (1992) attempted 
to correct for this by adjusting the surface pressure boundary conditions to 
force agreement with upper giant branch sequences for globular clusters with 
infrared observations. However, it is evident from the M13 and M92 sequences on 
Figure 6 of VandenBerg (1992) that the corrected models still underestimate the 
metallicities of metal-poor clusters. VandenBerg (1992) noted that there 
are a number of possible causes for this effect. 
The LFs predicted from these models also do not match those of 
very metal-poor clusters near the RGB-tip (VandenBerg, Bolte, \& Stetson 1996), 
while the terminal RGB core masses are lower than those predicted by other 
models (Caloi, D'Antona, \& Mazzitelli 1997, Silvestri et al. 1998).

	The metallicities and color excesses derived for the inner spheroid 
clusters are listed in Table 6, along with the corresponding values listed by 
Harris (1996). The metallicities and reddenings listed for HP 1, Djorgovski 1, 
NGC 6139, and NGC 6287 supersede those given by Davidge (1998, 2000). 
The $B-V$ color excesses derived here are, on average, in good 
agreement with the Harris (1996) values, with a mean difference 
$\Delta E(B-V) = -0.06 \pm 0.04$, in the sense this study minus 
Harris. When corrected for the systematic offset in metallicity noted above, 
the metallicities of the inner spheroid clusters are 
in good agreement with those given by Harris, with $\Delta$[Fe/H] 
$= -0.07 \pm 0.06$. The correlation coefficient between the 
metallicities listed in the second and fifth columns of Table 6 is 0.62, which 
is significant at more than the 99\% confidence level. 

	The mean CO index on the giant branch, CO$^{RGB}$, which was measured 
by computing the mode of the CO distribution from the $(K, CO)$ CMD of each 
cluster, is sensitive to chemical composition. CO$^{RGB}$ values are listed 
in the middle column of Table 7; the last column in this table shows 
CO$^{RGB}$ corrected for reddening. There is significant scatter in the 
CO$^{RGB}$ measurements, with no obvious trend between CO$^{RGB}$ and 
[Fe/H]. The absence of a trend with metallicity is not surprising since only a 
modest variation in CO index is expected when [Fe/H] $\leq -1$. Bell \& Briley 
(1991) modelled CO indices in metal-poor giants, and predicted that the CO 
index should change by only $\sim 0.04$ mag between [M/H] $= -2$ and $-1$ if 
[C/Fe] is constant, which is comparable to the uncertainty in an individual 
observation. However, at a fixed metallicity the scatter in CO$^{RGB}$ is 
slightly larger than the uncertainties in the photometry. For the 
clusters with [Fe/H] $\leq -2$ the mean CO index is 0.00 with a standard 
deviation $\pm 0.06$, while for clusters with [Fe/H] $\geq -2$ the mean is
0.02 with $\sigma = \pm 0.07$ mag.

	Cluster-to-cluster variations in chemical composition may 
explain much of the scatter in CO$^{RGB}$. The Bell \& Briley (1991) 
models predict that the excess scatter in CO$^{RGB}$ would require [CNO/Fe] to 
vary by a few tenths of a dex, and an abundance dispersion of this size is not 
without precedent. Smith et al. (1996) measured [C/Fe] for stars in M13, 
and found two populations, with [C/Fe] $\sim -0.8$ and $-1.2$; an even larger 
difference is seen among [O/Fe] and [N/Fe] measurements for stars in this 
cluster (Kraft et al. 1992). Cluster-to-cluster variations in [C/Fe] may be 
smaller than the star-to-star variations within clusters, although the sample 
size is modest. The mean [C/Fe] values in M13 and M3 are $-1.0 \pm 0.1$ and 
$-0.9 \pm 0.1$, respectively, while the mean [C/Fe] in M10, which has a mean 
metallicity similar to M13 and M3, appears to be between --1.1 and --1.2 
based on spectra of two stars (Kraft et al. 1995). The mean [O/Fe] values 
of M13 and M3 differ by $\sim 0.2$ dex (Kraft et al. 1992).

	The abundance anomalies among giants in M13 and M3 are likely due to a 
combination of primordial and evolutionary effects (Kraft et al. 1992). Studies 
of halo dwarfs provide additional insight into primordial abundance variations, 
and surveys of these objects suggest that there is a dispersion in [C/Fe] of a 
few tenths of a dex (e.g. Carbon et al. 1987), although for [O/Fe] the scatter 
may be smaller (e.g. Boesgaard et al. 1999). It thus appears that the 
abundance dispersion observed among metal-poor giants and dwarfs 
may contribute significantly to the scatter in CO$^{RGB}$. 
Spectra will provide the definitive means of determining the size 
of cluster-to-cluster [CNO/Fe] variations among inner spheroid clusters.

\subsection{Distances}

	The reddening-corrected RGB-tip measurements are listed in the second 
column of Table 8. The histogram distribution of these values, shown in the 
lower panel of Figure 10, is symmetric, indicating that the cluster sample is 
not biased towards objects on one side of the GC. 

	Distances to individual clusters were computed using two different 
M$_K^{RGBT}$ calibrations. The Bergbusch \& VandenBerg (1992) models, when 
transformed onto the infrared observational plane, predict that M$_K^{RGBT} = 
-5.9$ for metal-poor clusters, with only a modest metallicity dependence when 
[Fe/H] $\leq -1.3$. The reddening-corrected distance moduli predicted from this 
calibration, $\mu^{BV}$, are listed in the third column of Table 8. If it is 
assumed that the clusters are distributed isotropically, then these 
entries can be used to estimate the distance to the GC after correcting 
for the geometric displacement from the GC sight line. This calibration 
gives a distance to the GC of $15.1 \pm 0.1$, which is markedly higher than 
that computed by Reid (1993).

	Cluster distances were also computed using an empirical M$_K^{RGBT}$ 
calibration based on near-infrared observations made by Cohen et al. (1978) and 
Frogel et al. (1983) of bright giants in halo clusters with [Fe/H] 
between --1.5 and --2.0 and $E(B-V) \leq 0.1$. The clusters that 
meet these selection criteria are NGC 5272, NGC 5897, NGC 6205, and NGC 6752, 
and the distances of these clusters were calculated using HB brightnesses from 
Harris (1996) and the HB calibration of Carney et al. (1992). 
These data predict that M$_K^{RGBT} = -5.7$, and 
the distance moduli predicted with this calibration, $\mu_{emp}$, 
are listed in the fourth column of Table 8. 
This calibration gives a distance to the GC of $14.9 \pm 0.1$, 
which is in slightly better agreement with the Reid (1993) value.

	Walker (1992) used RR Lyrae variables in the LMC to set the zeropoint 
for the RR Lyrae brightness calibration, and found a 0.3 mag offset with 
respect to the Carney et al. (1992) zeropoint. If the Walker (1992) RR Lyrae 
calibration had been adopted to determine the empirical M$_K^{RGBT}$ 
calibration in the preceding paragraph then the distance moduli listed in 
the fourth column of Table 8 would increase by 0.3 mag. The Walker (1992) 
calibration would thus give a distance to the GC in better agreement 
with the RGB-tip calibration from the Bergbusch \& VandenBerg models, but 
significantly different from that estimated by Reid (1993). 

	The distance moduli predicted from the HB brightnesses, metallicities, 
and reddenings listed by Harris (1996), assuming the Carney et al. (1992) 
HB calibration, are listed in the fifth column of Table 8. The 
histogram distribution of the difference between $\mu^{emp}$ and $\mu^{H96}$, 
$\Delta \mu$, is shown in Figure 11, and there is an offset of $\sim 0.3$ 
dex between the two sets of distance estimates.

	There are three clusters with large $\Delta \mu$ that define the 
tail of the distribution in Figure 11: NGC 6293, NGC 6558, 
and NGC 6642. All three clusters have $C < 10$, and hence are subject to 
extreme bulge star contamination. Moreover, the $\mu^{H96}$ values for NGC 
6642 and NGC 6558 are very uncertain. NGC 6642 does not have a published 
CMD at visible wavelengths, and the RR Lyrae brightness listed by 
Harris (1996) is based on a survey for variable stars 
by Hazen (1993). The scatter in the mean RR Lyrae brightnesses within 
the tidal radius of NGC 6642 is comparable to that outside the tidal radius, 
suggesting that some of the variables identified by Hazen (1993) as possible 
cluster members may instead belong to the field. As for NGC 6558, the 
$(V, V-I)$ CMD for this cluster published by Rich et al. 
(1998), which is the source of the HB brightness measurement in the most recent 
version of the Harris database, shows considerable scatter, with a poorly 
defined RGB and HB. The situation is different for NGC 6293, as the $(V, B-V)$ 
CMD of this cluster obtained by Janes \& Heasley (1991) shows a moderately 
well-defined giant branch and HB; hence, $\mu^{H96}$ for this cluster 
is likely reliable.

	NGC 6293 and NGC 6558 are also core-collapsed clusters (Trager, King, 
\& Djorgovski 1995), and stellar content may change with radius in objects of 
this nature, including a central depletion of upper giant branch 
stars (e.g. Djorgovski et al. 1991, Djorgovski \& Piotto 1993). The central 
depletion of bright giants appears to be a common phenomenon in clusters 
(e.g. Shara et al 1998), including 47 Tuc (Bailyn 1994); nevertheless, the 
effect appears to be greatest in core-collapsed clusters, with NGC 7099 being 
one of the most extreme examples (Davidge 1995; Burgarella \& Buat 1996). A 
physical process that can deplete the central population of bright giants while 
also matching the statistics of other stars has not been identified (e.g. 
Djorgovski et al. 1991).

	It is likely that the bright giant content will be depleted in some of 
the inner cluster fields considered in the current sample; indeed, the CMD of 
NGC 6293 presented by Janes \& Heasley (1991) reveals a dearth of stars on the 
upper giant branch, and this is likely why this cluster has a large $\Delta 
\mu$. However, NGC 6293 appears to be an extreme example of upper giant branch 
depletion, and in the majority of clusters giant branch depletion occurs only 
within a few arcsec of the cluster center (e.g. Djorgovski \& Piotto 1993). 
The inner cluster fields sample significantly larger areas than this, so it can 
be anticipated that the distances derived for the majority of inner spheroid 
clusters will not be affected by the depeletion of giants. To investigate if 
this is the case, the distances to clusters that Trager et al. (1995) found 
to be core-collapsed were compared with those that are not core-collapsed. The 
structural nature of each cluster is given in the last column of Table 8, and 
a question mark identifies clusters with uncertain central 
morphologies. Djorgovski 1 was not examined by Trager et al. (1995).

	The mean $\Delta \mu$ for the eight confirmed core-collapsed clusters 
is $0.5 \pm 0.3$, while for the eight clusters that are not core-collapsed the 
mean is $= 0.3 \pm 0.1$. Hence, the mean $\Delta \mu$ values for the two groups 
of clusters are not significantly different. The histogram distribution of 
$\Delta \mu$ values for core-collapsed cluster is shown in Figure 11, and 
it is evident that while two of the three clusters with the largest $\Delta 
\mu$ are core-collapsed, the majority of core-collapsed 
systems have $\Delta \mu$ similar to the majority of other clusters. 
Therefore, while some clusters may have depleted upper giant branch 
populations, with NGC 6293 being the most extreme example, the majority of 
core-collapsed clusters in this sample have upper giant branches that are 
sufficiently well populated to allow the RGB-tip to be identified.
	 
	The $\mu^{emp}$ entries in Table 8 span a range of values, with a 
significant fraction of the clusters located on the far side of the bulge. 
The spatial distribution of the metal-poor inner spheroid cluster sample is 
thus very different from that for the metal-rich clusters studied by 
Barbuy, Bica, \& Ortolani (1998), which were found to fall exclusively 
on the near side of the bulge.

\subsection{The Spectral-Energy Distributions of Giant Branch Stars}

	Davidge \& Courteau (1999a) found that giants in NGC 6287 may not 
fall along the same sequence as stars in metal-poor halo clusters in the $(J-H, 
H-K)$ TCD. Do giants in the larger sample of metal-poor 
inner spheroid clusters support this finding?
To answer this question, the giant branch sequences of all twenty inner 
spheroid clusters were placed on the near-infrared TCD and compared with the 
halo cluster sequence. To reduce scatter, this comparison was done 
using normal points for the inner spheroid clusters, which were computed by 
finding the mode of the color distribution in $\pm 0.25$ mag bins along the $K$ 
axes of the $(K, H-K)$ and $(K, J-K)$ CMDs, and the resulting TCD is shown in 
Figure 12.

	The data for the inner spheroid clusters is distributed over 0.12 mag 
along the $H-K$ axis, which is consistent with the scatter in the standard 
star observations. Therefore, there is no evidence 
for an intrinsic dispersion in the near-infrared SEDs of inner spheroid 
clusters. In addition, the midpoint of the inner spheroid cluster data 
falls 0.02 mag to the right of the halo cluster sequence in Figure 12, which is 
smaller than the uncertainties in the photometric zeropoints. These data thus 
indicate that the near-infrared SEDs of metal-poor inner spheroid clusters are 
not significantly different from those of metal-poor halo clusters.

\subsection{Integrated Spectral-Energy Distributions}

	The $(J-H, H-K)$ and $(CO, J-K)$ TCDs constructed from the integrated 
color measurements of the inner spheroid clusters are shown in Figure 13. While 
it might be anticipated that dynamical evolution will influence the integrated 
SEDs of clusters, Figure 13 shows that the integrated colors of core-collapsed 
clusters are not vastly different from those of other clusters in this sample.

	The $(CO, J-K)$ TCD in Figure 13 suggests that CO weakens as J-K 
increases, which is contracy to what would be expected if the SED and CO index 
is defined by a single parameter: metallicity. As demonstrated below, 
the trend in this TCD is due to a small number of clusters that 
have peculiar integrated CO indices.

	Aaronson et al. (1978) investigated the 
near-infrared color -- metallicity relations for globular clusters, and these 
have subsequently been re-calibrated by Davidge (2000) using more 
up-to-date metallicities. The integrated $J-K$ and CO colors of the 
inner spheroid clusters, corrected for reddening using the entries in the 
fourth column of Table 6, are plotted as functions of metallicity in 
the upper panels of Figure 14.

	The majority of clusters fall along the halo cluster $(J-K) -$ [Fe/H] 
relation, and the scatter envelope defined by
core-collapsed clusters is not greatly different than that defined by 
other clusters. The [Fe/H] $= -1.9$ clusters NGC 6144 and 
NGC 6293 have $(J-K)_0$ colors that are too large 
for their metallicities. The very red color of NGC 6293 is somewhat 
surprising since, as a core-collapsed cluster with a depleted upper giant 
branch (Janes \& Heasley 1991), this object might 
be expected to fall below the $(J-K)_0$ versus [Fe/H] relation. 
The metallicities computed for both clusters are in reasonable agreement with 
those given by Harris (1996), so errors in metallicity do not present an 
obvious explanation for the red colors. The presence of bulge stars 
can affect integrated color measurements (Davidge 2000), although the $(K, CO)$ 
CMDs of these clusters indicate that a large population of bright bulge stars 
is not present in the inner cluster fields, where the integrated photometric 
measurements were made. The reddenings derived for these clusters are both 
smaller than those listed by Harris (1996), and if the reddenings have been 
underestimated then this could explain the position of these clusters in 
the upper panel of Figure 14. Independent photometric observations 
will resolve this issue.

	A significant number of inner spheroid clusters depart from the halo 
cluster CO$_0 -$ [Fe/H] relation. Core-collapsed clusters show the largest 
departures, although other clusters in the sample show marked deviations as 
well. The majority of clusters that depart from the halo cluster 
relation are the most metal-poor in this sample, and the scatter is too large 
to be due to uncertainties in metallicity and/or reddening. The 
relative behaviours of the integrated CO and CO$^{RGB}$ indices is investigated 
in the lower panel of Figure 14. It is evident that while these quantities 
are related for the majority of clusters, there are some clusters (NGC 6284, 
NGC 6287, NGC 6293, NGC 6333, and NGC 6558) that fall off the relation, as they 
have very small integrated CO indices; the CO${RGB}$ indices for these same 
clusters are not different from those of the other clusters, so a problem with 
the photometric calibration can not explain the very small integrated CO 
indices. These data thus suggest that the integrated CO index in some 
metal-poor inner spheroid clusters may be affected by quantities other than 
metallicity. 

\section{THE LUMINOSITY FUNCTION OF THE BULGE}

	It might be anticipated that the reddenings of some of the cluster and 
nearby bulge fields may differ due to the patchy nature of obscuration 
at low Galactic latitudes. Unfortunately, the low stellar densities in many of 
the bulge fields hinder efforts to measure reddening directly. However, the 
reddenings estimated with the procedure described by Davidge (2000), which 
assumes that the colors of the brightest stars are similar 
to those of M giants in Baade's Window (BW), 
for those fields with relatively high stellar densities 
are not significantly different from those of the nearby metal-poor clusters. 
Therefore, for the subsequent analysis the cluster reddenings are adopted 
for the bulge fields. 

	The Field 3 LFs are compared in Figure 15, where the $E(B-V)$ entries 
in column 4 of Table 6 have been used to correct stellar brightnesses for 
extinction. The LFs follow power-laws near the bright end, and typically extend 
to within a magnitude of the bulge MSTO, which occurs near K$_0 = 18$ if 
M$_K \sim 3$ (e.g. Bertelli et al. 1994). The onset of the SGB is seen when 
$K_0 \geq 16$ in some fields as a departure from the power-law defined by 
brighter stars. A bump due to the bulge HB is also evident near $K = 13$ in the 
LFs of HP 1 Field 3, NGC 6453 Field 3, and NGC 6558 Field 3. 

	The rate of evolution on the upper giant branch is insensitive to 
metallicity (VandenBerg 1992), so the power-law exponent that characterizes the 
bright ends of the LFs should not vary significantly among the Field 3 
datasets, even though the mean metallicity of the bulge changes with radius 
(e.g. Minniti et al. 1995b). Exponents were measured for each Field 3 LF by 
using the method of least squares to fit a power-law between K$_0 = 14$ 
and 15.5, and the results are listed in Table 9. The fitting procedure was 
restricted to this particular brightness interval because (1) it avoids the 
bulge HB, (2) a reasonable number of stars are detected over this range of 
brightnesses in all 18 bulge fields, and (3) with the exception of HP 1 
Field 3, the data are complete to $K_0 = 15.5$ in all fields. 

	A reference LF was constructed by co-adding the de-reddened Field 3 
LFs, and the result has an exponent $x = 0.335 \pm 0.018$. The HP 1 Field 3 
observations were not included when constructing this reference LF because 
the effects of crowding become significant at relatively bright levels in this 
field. In fact, it is evident from Figure 15 that the LF of HP 1 Field 3 
at the bright end is much stepper than at the faint end, 
suggesting that incompleteness becomes significant when $K_0 \geq 14$.

	DePoy et al. (1993) conducted a 
wide-field survey of bright giants in BW, including the region around NGC 6522, 
and a least squares fit to the data listed in Table 2 of that study gives 
an exponent $x = 0.275 \pm 0.005$, which is in excellent agreement with the 
LF exponent for NGC 6522 Field 3 listed in Table 9. The LF exponents for NGC 
6273 and NGC 6522 Field 3 differ from that of the co-added LF at roughly the 
$2.5\sigma$ level, and so these data suggest 
that the slope of the LF may vary throughout the bulge. 

\section{DISCUSSION \& SUMMARY}

\subsection{Cluster Properties}

	Images recorded through $J, H, K, 2.2\mu$m continuum, and $2.3\mu$m CO 
filters have been used to estimate the reddenings, metallicities, 
and distances of the clusters with [Fe/H] $\leq -1.3$ and R$_{GC} \leq 
3$ kpc listed in the May 1996 version of the Harris (1996) database. The 
clusters are located in crowded bulge fields, and tend to be highly reddened. 
Metallicities and reddenings are estimated from the shape and color of the 
upper giant branch on the $(K, J-K)$ CMD, while distance is determined from the 
$K$ brightness of the RGB-tip. These data provide an independent 
means of determining basic cluster properties at wavelengths where 
(1) the effects of extinction are greatly reduced with respect to the visible, 
and (2) there are strong absorption features, such as the $2.3\mu$m CO 
bands, that can be used to identify bright bulge stars. 
The metallicities and reddenings derived from the near-infrared 
CMDs tend to be in good agreement with those given by Harris (1996). 

	The survey is not a complete census of metal-poor 
inner spheroid clusters, as uncertainties in the published values of 
[Fe/H] and R$_{GC}$, on which the sample selection is based, have 
likely caused clusters with intrinsic parameters matching the selection 
criteria to be excluded. There are also clusters that remain to be discovered, 
especially between Galactic latitudes --2 and 1 degrees (Barbuy, 
Bica, \& Ortolani 1998). While the clusters closest to the GC are of the 
greatest astrophysical interest, they are also likely rare, 
given the low density of clusters in the central regions of 
nearby spiral galaxies (e.g. Battistini et al. 1993; Davidge \& Courteau 1999b).

	The identification of bulge stars, which is a critical part of 
any survey of globular clusters at low Galactic latitudes, is best 
done using spectroscopic information, such as the depth of the 
$2.3\mu$m CO bands. If not identified and removed 
from the data, bright bulge stars can significantly bias calculated 
cluster properties. For example, the brightest bulge stars are on the near side 
of the bulge and, if these are mistakenly assumed to be cluster members, 
they will skew cluster distances to values less than that of the GC. 
Bulge star contamination could thus explain why Barbuy et al. 
(1998) find an absence of metal-rich clusters on the far side of the bulge. 
Indeed, the current data suggests that Djorgovski 1, which is 
in a densely populated bulge field and is one 
of the clusters in the Barbuy et al. sample, 
is on the far side of the Galaxy. In addition, the 
majority of bulge stars have [Fe/H] $\geq -1$, and the inclusion of these 
objects in cluster studies will skew metallicities upwards. This effect 
has been well documented in the particular case of HP 1 (e.g. Ortolani, Bica, 
\& Barbuy 1997, Bica et al. 1998, Davidge 2000). 

	The $K$ brightness of the RGB-tip, which is determined from bright 
stars that can be identified as cluster members using CO indices, is a useful 
supplemental distance indicator for globular clusters, especially for those 
that are heavily reddened and occur in dense bulge fields, where the HB 
brightness may be highly uncertain. In fact, a small sample of clusters in this 
sample have distances that are significantly different from those computed from 
the HB brightnesses listed by Harris (1996), with the greatest discrepancy 
occuring for NGC 6558. This cluster is located in a richly populated bulge 
field, and contamination from bulge stars is significant, even within a few 
tens of arcsec of the cluster center. NGC 6558 is also core-collapsed; however, 
the majority of core-collapsed clusters in this sample have unremarkable (when 
compared with other inner spheroid cluster) LF 
exponents, and RGB-tip distances that are similar to non-collapsed clusters. 
Thus, it appears that the upper giant branches of the majority of 
core-collapsed clusters in the current sample have not been dramatically 
altered by dynamical evolution, at least over the angular scales sampled with 
the current data (i.e. a few tens of arcsec from the cluster center). A 
significant exception is NGC 6293, which has a depleted upper giant branch 
population over a large part of the cluster (Janes \& Heasley 1991).

\subsection{Comparing the Observational Properties of Metal-Poor Clusters in 
the Inner and Outer Spheroid}

	HB morphology provides a natural observational tool for comparing 
the properties of metal-poor clusters. Early observations suggested that 
the HBs of globular clusters in the inner regions of the Galaxy 
were influenced by only a single parameter -- metallicity (e.g. Searle 
\& Zinn 1978, Lee 1992). If age is the only other parameter 
influencing HB morphology then such a uniformity in HB content 
indicates that the inner regions of the Galaxy experienced a rapid collapse, 
whereas clusters in the outer halo formed over an extended period of time 
(Searle \& Zinn 1978, Sarajedini, Chaboyer, \& Demarque 1997, Buonanno et al. 
1998). However, this picture has recently been challenged in two ways. 
First, deep photometric studies of metal-poor clusters within a few kpc 
of the GC, such as NGC 6287 (Davidge \& Courteau 1999a, Stetson \& West 1994), 
Terzan 1 (Ortolani et al. 1999), and NGC 6558 (Rich et al. 1998) reveal 
HB morphologies that differ from those of old halo clusters. 
Second, parameters other than age and metallicity may 
influence HB morphology (e.g. Salaris \& Weiss 1997, Buonanno et al. 1997), 
thereby complicating the interpretation of HB data. Consequently, it is 
of interest to investigate other ways in which the observational 
properties of inner and outer spheroid clusters differ. 

	The current data indicate that inner spheroid clusters tend to 
have steeper giant branch LFs than halo clusters. Clusters in the 
inner and outer spheroid have experienced different dynamical histories (e.g. 
Murali \& Weinberg 1997, Vesperini 1997, 1998, and references therein), and 
this could affect the bright stellar content. The appearance of 
population gradients within clusters is related to dynamical evolution 
(e.g. discussion in \S 4.2), and dynamical evolution 
will likely occur at a rapid pace in the densely-populated inner regions of the 
Galaxy. Indeed, Rich et al. (1998) suggest that NGC 6522, NGC 6558, and 
HP 1 may be part of a new class of dynamically evolved clusters, characterised 
by a larger than average ratio of HB stars to bright giants, that are the 
result of interactions in the inner Galaxy.

	If the stellar contents of inner spheroid clusters have been affected 
by dynamical interactions then this will influence integrated photometric 
properties. However, broad-band near-infrared colors are relatively insensitive 
to changes in mass function exponent. Vesperini \& Heggie (1997) conclude that 
the present-day mass function in all but the most massive clusters with R$_{GC} 
\leq 3$ kpc differs significantly from the initial relation, and the 
main sequence mass function exponent might change by $\Delta x = 2$ 
for clusters with small R$_{GC}$. Buzzoni (1989) modelled the broad-band colors 
of simple stellar systems for a range of mass function exponents, and his 
models indicate that changes of the order $\Delta x = 2$ will alter $J-K$ by 
only 0.05 mag. The Buzzoni (1989) models do not consider, for example, merged 
stellar objects, the central depletion of bright giants, or mass segregation, 
and so they are not well-suited for studying the impact of dynamical evolution 
on photometric properties; nevertheless, these simple models still indicate 
that large changes in stellar content are required to alter integrated infrared 
colors by $\sim 0.1$ mag. 

	Many inner spheroid clusters depart from 
the integrated CO versus [Fe/H] relation defined by halo 
clusters, and this result is difficult to explain. A centrally enhanced 
blue population, such as that found by Rich et al. (1998) in the so-called 
NGC 6522 class of clusters, would decrease the strength of $2.3\mu$m CO 
absorption. However, two of the members of this proposed class, NGC 6522 and 
HP 1, fall along the CO versus [Fe/H] relation defined by halo clusters. In any 
event, the presence of a blue population that is large enough to affect the 
CO indices would also affect broad-band colors, and the $J-K$ colors of 
the clusters in the current sample are normal for their metallicity. The 
departures from the integrated CO versus metallicity relation are also 
likely not due to extreme [CNO/Fe] abundances, as the scatter in the 
CO$^{RGB}$ indices indicate that any cluster-to-cluster differences 
in [CNO/Fe] are not much larger than what has already been detected 
among halo stars.

\subsection{Bright Giants and HB Stars in the Bulge}

	The bulge fields considered in this study demonstrate the 
heterogeneous nature of the inner spheroid. The LFs of NGC 6273 Field 3 and 
BW, which is sampled by NGC 6522 Field 3, are significantly shallower than the 
composite bulge LF. Consequently, BW appears not to be a `typical' inner bulge 
field in terms of the statistics of upper giant branch stars. The uncertainties 
in the LF exponents measured for some fields are large, and wide-field 
near-infrared photometric surveys of the bulge will produce datasets from which 
more reliable exponents can be measured, especially at large R$_{GC}$.

	The slope of the upper giant branch on CMDs can be used to 
estimate mean metallicity. Tiede, Frogel, \& Terndrup (1995) measured the 
giant branch slope in a number of bulge fields, and 
used this information to study the bulge metallicity gradient. 
Unfortunately, while the bulge fields observed for the current study sample a 
significant range of environments, in many cases the upper giant branch 
is not well populated. There are 7 bulge fields with moderately 
large numbers of stars brighter than K$_0 = 12.6$, which is the brightness 
interval considered by Tied et al., and the slopes measured 
from the $(K, J-K)$ CMDs of these fields are listed in Table 10; the fields 
in this Table are ordered according to distance above the Galactic Plane. 
The slope of the giant branch in NGC 6522 Field 3 is in excellent 
agreement with that measured by Tiede et al. (1995) for BW.

	The slopes listed in Table 10 show a great deal of scatter, with no 
obvious trend with $\mid b \mid$. The uncertainties in the 
slope estimates propogate into significant errors in [Fe/H] estimates, 
which were computed from Equation 1 of Tiede et al. (1995) and are 
listed in the fourth column of Table 10.

	The fields considered in Table 10 do not cover a large enough range of 
$\mid b \mid$ to permit an independent investigation of the bulge metallicity 
gradient. However, the metallicities derived here can be combined with those 
measured in other studies. The metallicities for NGC 6522 Field 3 and NGC 6453 
Field 3 were averaged to create a single point for $\mid b \mid$ = 
3.9, while the metallicities for the last 
four points were averaged to produce a mean for $\mid b \mid$ = 6. These 
were then combined with the data in Table 8 of Tiede et al. (1995) to create 
new average metallicities at these $\mid b \mid$. The results for HP 1 
Field 3 were also combined with the Terzan 2 entry in that table. 
A least squares fit was then applied to the data in the last column 
of Table 8 of Tiede et al. (1995), with the entries for $\mid b \mid$ = 
2.3, 3.9, and 6 adjusted as described above, and the gradient measured in 
this manner is $\Delta$[Fe/H]/$\Delta \mid b \mid = -0.068 \pm 0.016$.

	The bulge data are also suggestive of 
field-to-field differences in HB content. In particular, while 
the HB is not conspicuous in the majority of Field 3 LFs in Figure 15, 
a HB bump is clearly evident near $K_0 = 13.5$ in the NGC 6453 and NGC 6558 
background field LFs. There is also a tantalizing hint of a HB bump 
in the HP 1 Field 3 LF, but the effects of incompleteness (\S 5) make the 
reality of this feature difficult to quantify with the existing data. The NGC 
6453 and NGC 6558 background fields sample regions of the bulge that are 
roughly 6 degrees from the GC, suggesting a possible connection with distance 
from the GC. However, NGC 6355 Field 3 samples a region that is 5.4 degrees 
from the GC, and a distinct HB is not evident in the LF of this field; 
consequently, the current data are not consistent with a simple radial trend.

	The $K$ LFs of NGC 6453 Field 3, NGC 6558 Field 3, and NGC 6355 Field 3 
in a narrow brightness interval centered on the HB are compared in Figure 16; 
the composite bulge LF described in \S 5 is also plotted in this figure. The 
LFs in this figure were normalized and then shifted by arbitrary amounts along 
the vertical axis for the purposes of display. The strong HB features in 
the NGC 6453 Field 3 and NGC 6558 Field 3 data are clearly evident. A 
Kolmogoroff-Smirnoff test indicates that the combined LFs of both fields 
between K = 12.5 and 14.0 differs from that of the composite bulge LF at 
roughly the $2\sigma$ level. 

	These data suggest that the HB content is not uniform throughout 
the bulge, and a moderately deep photometric near-infrared survey of fields 
within a few arcmin of the GC should be able to verify or refute this result. 
The uniformity of the R' statistic at moderate distances from the 
GC (Minniti 1995) suggests that variations in HB content, if present, will be 
restricted to the inner spheroid. The presence of variations 
in HB content are not entirely unexpected given that (1) 
the bulge contains a metallicity gradient, and (2) HB morphology changes with 
metallicity. Age will also affect HB content, and Paczynski et al. (1994) and 
Kiraga, Paczynski \& Stanek (1997) have noted that an intermediate-age 
population may explain the large number of red HB stars in BW. If the 
relative contribution of such an intermediate age component to total 
light varies with location in the bulge then this will also cause  
a non-uniform HB content.

\vspace{0.5cm}
	Sincere thanks are extended to Jim Hesser and Peter Stetson for 
commenting on an earlier version of this paper.

\clearpage 

\begin{table*}
\begin{center}
\begin{tabular}{lrrc}
\tableline\tableline
Cluster & $l_{II}$ & $b_{II}$ & Extraction \\
 & & & Radius (arcsec) \\
\tableline
Djorgovski 1 & 356.67 & --2.48 & 45 \\
HP 1 & 357.42 & 2.12 & 35 \\
NGC 6093 & 352.67 & 19.46 & -- \\
NGC 6139 & 342.37 & 6.94 & 50 \\
NGC 6144 & 351.93 & 15.70 & -- \\
NGC 6235 & 358.92 & 13.52 & 25 \\
NGC 6273 & 356.87 & 9.38 & -- \\
NGC 6284 & 358.35 & 9.94 & -- \\
NGC 6287 & 0.13 & 11.02 & 65 \\
NGC 6293 & 357.62 & 7.83 & -- \\
NGC 6333 & 5.54 & 10.70 & -- \\
NGC 6355 & 359.58 & 5.43 & 45 \\
NGC 6453 & 355.72 & --3.87 & 45 \\
NGC 6522 & 1.02 & --3.93 & 45 \\
NGC 6541 & 349.29 & --11.18 & -- \\
NGC 6558 & 0.20 & --6.03 & 45 \\
NGC 6626 & 7.80 & --5.58 & 45 \\
NGC 6642 & 9.81 & --6.44 & 35 \\
NGC 6681 & 2.85 & --12.51 & 35 \\
NGC 6717 & 12.88 & --10.90 & 40 \\
\tableline
\end{tabular}
\end{center}
\caption{METAL-POOR INNER SPHEROID CLUSTER SAMPLE}
\end{table*}

\clearpage

\begin{table*}
\begin{center}
\begin{tabular}{cc}
\hline\hline
Cluster & $x$ \\
\hline
NGC 288 & $0.22 \pm 0.03$ \\
NGC 362 & $0.15 \pm 0.02$ \\
NGC 7089 & $0.16 \pm 0.02$ \\
\hline
\end{tabular}
\end{center}
\caption{POWER-LAW EXPONENTS FOR HALO CLUSTER K LFs}
\end{table*}

\clearpage

\begin{table*}
\begin{center}
\begin{tabular}{lcr}
\hline\hline
Cluster & $x$ & $C$ \\
\hline
Djorgovski 1 & $0.33 \pm 0.08$ & 2 \\
HP 1 & $0.17 \pm 0.11$ & 4 \\
NGC 6093 & $0.23 \pm 0.05$ & 50 \\
NGC 6139 & $0.18 \pm 0.05$ & 153 \\
NGC 6144 & $0.31 \pm 0.12$ & 17 \\
NGC 6235 & $0.13 \pm 0.16$ & 15 \\
NGC 6273 & $0.23 \pm 0.02$  & 55 \\
NGC 6284 & $0.25 \pm 0.07$ & 7 \\
NGC 6287 & $0.38 \pm 0.02$ & 158 \\
NGC 6293 & $0.25 \pm 0.07$ & 4 \\
NGC 6333 & $0.20 \pm 0.08$ & 14 \\
NGC 6355 & $0.19 \pm 0.07$ & 6 \\
NGC 6453 & $0.25 \pm 0.12$ & 6 \\
NGC 6522 & $0.31 \pm 0.07$ & 3 \\
NGC 6541 & $0.27 \pm 0.04$ & 33 \\
NGC 6558 & $0.42 \pm 0.08$ & 3 \\
NGC 6626 & $0.23 \pm 0.09$ & 9 \\
NGC 6642 & $0.17 \pm 0.03$ & 7 \\
NGC 6681 & $0.13 \pm 0.11$ & 43 \\
NGC 6717 & $0.31 \pm 0.13$ & 5 \\
\hline
\end{tabular}
\end{center}
\caption{POWER-LAW EXPONENTS AND FIELD STAR CONTAMINATION INDICES FOR INNER 
SPHEROID CLUSTERS}
\end{table*}

\clearpage

\begin{table*}
\begin{center}
\begin{tabular}{lccccr}
\hline\hline
Cluster & $K_{RGBT}$ & $J-H$ & $H-K$ & $J-K$ & $CO$ \\
\hline
NGC 288 & 8.5 & -- & -- & -- & -- \\
NGC 362 & 8.2 & 0.46 & 0.09 & 0.55 & 0.10 \\
NGC 7089 & 9.7 & 0.68 & --0.04 & 0.64 & --0.04 \\
 & & & & & \\
Djorgovski 1 & 10.0 & 0.99 & 0.23 & 1.22 & 0.01 \\
HP 1 & 9.1 & 0.77 & 0.30 & 1.07 & -0.05 \\
NGC 6093 & 9.0 & 0.50 & 0.18 & 0.68 & --0.03 \\
NGC 6139 & 9.7 & 0.71 & 0.24 & 0.95 & --0.09 \\
NGC 6144 & 9.5 & 0.60 & 0.35 & 0.95 & --0.09 \\
NGC 6235 & 10.0 & 0.56 & 0.19 & 0.75 & --0.06 \\
NGC 6273 & 8.5 & 0.56 & 0.22 & 0.78 & --0.02 \\
NGC 6284 & 9.7 & 0.66 & 0.28 & 0.94 & --0.27 \\
NGC 6287 & 9.1 & 0.65 & 0.16 & 0.81 & --0.19 \\
NGC 6293 & 10.3 & 0.50 & 0.38 & 0.88 & --0.35 \\
NGC 6333 & 9.5 & 0.70 & 0.28 & 0.98 & --0.26 \\
NGC 6355 & 9.5 & 0.85 & 0.31 & 1.16 & 0.12 \\
NGC 6453 & 9.4 & 0.70 & 0.24 & 0.94 & 0.10 \\
NGC 6522 & 8.9 & 0.65 & 0.22 & 0.87 & --0.04 \\
NGC 6541 & 8.8 & 0.53 & 0.09 & 0.62 & --0.04 \\
NGC 6558 & 10.2 & 0.72 & 0.26 & 0.98 & --0.21 \\
NGC 6626 & 8.2 & 0.68 & 0.19 & 0.87 & 0.02 \\
NGC 6642 & 10.0 & 0.67 & 0.12 & 0.79 & 0.10 \\
NGC 6681 & 8.7 & 0.49 & 0.07 & 0.56 & 0.03 \\
NGC 6717 & 9.0 & 0.63 & 0.12 & 0.75 & --0.06 \\
\hline
\end{tabular}
\end{center}
\caption{RGB-TIP BRIGHTNESSES \& INTEGRATED COLORS}
\end{table*}

\clearpage

\begin{table*}
\begin{center}
\begin{tabular}{lccccc}
\tableline\tableline
Cluster & [Fe/H]$_{CMD}$ \tablenotemark{a} & $E(J-K)_{CMD}$ \tablenotemark{b} & $E(B-V)_{CMD}$ \tablenotemark{c} & [Fe/H]$_{H96}$ \tablenotemark{d} & $E(B-V)_{H96}$ \tablenotemark{e} \\
\hline
NGC 288 & --1.78 & 0.09 & 0.18 & --1.24 & 0.03 \\
NGC 362 & --1.26 & 0.02 & 0.04 & --1.16 & 0.05 \\
NGC 7089 & --2.03 & 0.05 & 0.10 & --1.62 & 0.05 \\
 & & & & & \\
M13 & --2.03 & 0.05 & 0.10 & --1.54 & 0.02 \\
\tableline
\end{tabular}
\end{center}
\caption{METALLICITIES AND REDDENINGS FOR HALO CLUSTERS}
\tablenotetext{a}{Cluster metallicity derived from the $(K, J-K)$ CMD.}
\tablenotetext{b}{$J-K$ color excess derived from the $(K, J-K)$ CMD.}
\tablenotetext{c}{$B-V$ color excess derived from E(J--K).}
\tablenotetext{d}{Metallicity listed by Harris (1996).}
\tablenotetext{e}{E(B--V) listed by Harris (1996).}
\end{table*}

\clearpage

\begin{table*}
\begin{center}
\begin{tabular}{lccccc}
\tableline\tableline
Cluster & [Fe/H]$_{CMD}$ \tablenotemark{a} & $E(J-K)_{CMD}$ \tablenotemark{b} & $E(B-V)_{CMD}$ \tablenotemark{c} & [Fe/H]$_{H96}$ \tablenotemark{d} & $E(B-V)_{H96}$ \tablenotemark{e} \\
\hline
Djorgovski 1 & --1.9 & 0.75 & 1.44 & -- & 1.7 \\
HP 1 & --1.3 & 0.44 & 0.85 & --1.50 & 1.19 \\
NGC 6093 & --1.6 & 0.05 & 0.10 & --1.62 & 0.18 \\
NGC 6139 & --1.9 & 0.33 & 0.63 & --1.65 & 0.74 \\
NGC 6144 & --1.9 & 0.11 & 0.21 & --1.73 & 0.32 \\
NGC 6235 & --1.4 & 0.17 & 0.33 & --1.40 & 0.36 \\
NGC 6273 & --1.9 & 0.20 & 0.38 & --1.68 & 0.37 \\
NGC 6284 & --1.3 & 0.15 & 0.29 & --1.32 & 0.28 \\
NGC 6287 & --1.9 & 0.22 & 0.42 & --2.05 & 0.59 \\
NGC 6293 & --1.9 & 0.02 & 0.04 & --1.92 & 0.39 \\
NGC 6333 & --1.9 & 0.27 & 0.52 & --1.72 & 0.36 \\
NGC 6355 & --1.6 & 0.56 & 1.08 & --1.50 & 0.75 \\
NGC 6453 & --1.9 & 0.33 & 0.63 & --1.53 & 0.61 \\
NGC 6522 & --1.4 & 0.29 & 0.56 & --1.52 & 0.50 \\
NGC 6541 & --1.9 & --0.08 & --0.15 & --1.83 & 0.12 \\
NGC 6558 & --1.9 & 0.23 & 0.44 & --1.44 & 0.42 \\
NGC 6626 & --1.3 & 0.24 & 0.46 & --1.45 & 0.41 \\
NGC 6642 & --1.9 & 0.15 & 0.29 & --1.35 & 0.40 \\
NGC 6681 & --1.1 & 0.02 & 0.04 & --1.51 & 0.07 \\
NGC 6717 & --0.9 & 0.06 & 0.12 & --1.29 & 0.21 \\
\tableline
\end{tabular}
\end{center}
\caption{METALLICITIES AND REDDENINGS FOR INNER SPHEROID CLUSTERS}
\tablenotetext{a}{Cluster metallicity derived from the $(K, J-K)$ CMD, and 
corrected for the systematic offset discussed in the text.}
\tablenotetext{b}{$J-K$ color excess derived from the $(K, J-K)$ CMD.}
\tablenotetext{c}{$B-V$ color excess derived from E(J--K).}
\tablenotetext{d}{Metallicity listed by Harris (1996).}
\tablenotetext{e}{E(B--V) listed by Harris (1996).}
\end{table*}

\clearpage

\begin{table*}
\begin{center}
\begin{tabular}{lrr}
\tableline\tableline
Cluster & CO$^{RGB}$ & CO$^{RGB}_0$ \\
\tableline
Djorgovski 1 & --0.04 & 0.02 \\
HP 1 & --0.01 & 0.02 \\
NGC 6093 & --0.01 & --0.01 \\
NGC 6139 & --0.02 & 0.01 \\
NGC 6144 & --0.08 & --0.07 \\
NGC 6235 & --0.05 & --0.04 \\
NGC 6273 & 0.01 & 0.03 \\
NGC 6284 & 0.12 & 0.13 \\
NGC 6287 & 0.08 & 0.10 \\
NGC 6293 & --0.06 & --0.06 \\
NGC 6333 & 0.01 & 0.03 \\
NGC 6355 & 0.06 & 0.10 \\
NGC 6453 & --0.10 & --0.07 \\
NGC 6522 & 0.04 & 0.06 \\
NGC 6541 & --0.04 & --0.04 \\
NGC 6558 & --0.09 & --0.07 \\
NGC 6626 & 0.04 & 0.06 \\
NGC 6642 & 0.04 & 0.05 \\
NGC 6681 & 0.00 & 0.00 \\
NGC 6717 & --0.10 & --0.10 \\
\tableline
\end{tabular}
\end{center}
\caption{GIANT BRANCH CO INDICES}
\end{table*}

\clearpage

\begin{table*}
\begin{center}
\begin{tabular}{lccccc}
\tableline\tableline
Cluster & K$_0$ & $\mu_0^{BV}$ \tablenotemark{a} & $\mu_0^{emp}$ \tablenotemark{b} & $\mu_0^{H96}$ \tablenotemark{c} & Core \\
 & & & & & Collapsed? \\
\tableline
Djorgovski 1 & 9.5 & 15.4 & 15.2 & -- & -- \\
HP 1 & 8.8 & 14.7 & 14.5 & 14.1 & Y \\
NGC 6093 & 9.0 & 14.9 & 14.7 & 14.5 & N \\
NGC 6139 & 9.5 & 15.4 & 15.2 & 14.9 & N \\
NGC 6144 & 9.4 & 15.3 & 15.1 & 14.9 & N \\
NGC 6235 & 9.9 & 15.8 & 15.6 & 14.8 & N \\
NGC 6273 & 8.4 & 14.3 & 14.1 & 14.5 & N \\
NGC 6284 & 9.6 & 15.5 & 15.3 & 15.6 & Y \\
NGC 6287 & 9.0 & 14.9 & 14.8 & 14.5 & N \\
NGC 6293 & 10.3 & 16.2 & 15.8 & 14.6 & Y \\
NGC 6333 & 9.3 & 15.2 & 15.0 & 14.4 & N \\
NGC 6355 & 9.1 & 15.0 & 14.8 & 14.1 & Y \\
NGC 6453 & 9.2 & 15.1 & 14.9 & 15.0 & Y \\
NGC 6522 & 8.7 & 14.6 & 14.4 & 14.1 & Y \\
NGC 6541 & 8.8 & 14.7 & 14.5 & 14.2 & ? \\
NGC 6558 & 10.0 & 15.9 & 15.7 & 13.9 & Y \\
NGC 6626 & 8.0 & 13.9 & 13.7 & 13.6 & N \\
NGC 6642 & 9.9 & 15.8 & 15.6 & 14.3 & ? \\
NGC 6681 & 8.7 & 14.6 & 14.4 & 14.5 & Y \\
NGC 6717 & 9.0 & 14.9 & 14.7 & 14.1 & ? \\
\tableline
\end{tabular}
\end{center}
\caption{CLUSTER DISTANCE MODULI}
\tablenotetext{a}{Distance modulus calculated from the RGB-tip using
the brightness calibration predicted by the Bergbusch \& VandenBerg (1992) 
models.}
\tablenotetext{b}{Distance modulus calculated from the RGB-tip calibrated using
observations of the brightest stars in metal-poor halo clusters and the 
Carney et al. (1992) RR Lyrae brightness calibration.}
\tablenotetext{c}{Distance modulus calculated from the HB brightnesses listed 
by Harris (1996) and the Carney et al. (1992) RR Lyrae calibration.}
\end{table*}

\clearpage

\begin{table*}
\begin{center}
\begin{tabular}{lc}
\tableline\tableline
Cluster & $x$ \\
\tableline
HP 1 & $0.070 \pm 0.004$ \\
NGC 6093 & $0.221 \pm 0.090$ \\
NGC 6139 & $0.476 \pm 0.115$ \\
NGC 6144 & $0.409 \pm 0.341$ \\
NGC 6235 & $0.237 \pm 0.086$ \\
NGC 6273 & $0.165 \pm 0.064$ \\
NGC 6284 & $0.672 \pm 0.225$ \\
NGC 6287 & $0.441 \pm 0.153$ \\
NGC 6293 & $0.386 \pm 0.079$ \\
NGC 6333 & $0.333 \pm 0.065$ \\
NGC 6355 & $0.390 \pm 0.081$ \\
NGC 6453 & $0.304 \pm 0.055$ \\
NGC 6522 & $0.246 \pm 0.032$ \\
NGC 6541 & $0.557 \pm 0.114$ \\
NGC 6558 & $0.377 \pm 0.032$ \\
NGC 6626 & $0.363 \pm 0.045$ \\
NGC 6642 & $0.352 \pm 0.071$ \\
NGC 6681 & $0.263 \pm 0.128$ \\
 & \\
Composite & $0.335 \pm 0.018$ \\
(No HP 1) & \\
\tableline
\end{tabular}
\end{center}
\caption{LUMINOSITY FUNCTION EXPONENTS MEASURED IN BULGE FIELDS}
\end{table*}

\clearpage

\begin{table*}
\begin{center}
\begin{tabular}{cccc}
\tableline\tableline
Field & $\mid b \mid$ & $\frac{\Delta K}{\Delta (J-K)}$ & [Fe/H] \\
\tableline
HP1 Field 3 & 2.1 & $-0.073 \pm 0.014 $ & $-1.24 \pm 0.33$ \\
NGC 6453 Field 3 & 3.9 & $-0.147 \pm 0.037$ & $+0.52 \pm 0.88$ \\
NGC 6522 Field 3 & 3.9 & $-0.105 \pm 0.013$ & $-0.48 \pm 0.31$ \\
NGC 6355 Field 3 & 5.4 & $-0.084 \pm 0.013$ & $-0.98 \pm 0.31$ \\
NGC 6626 Field 3 & 5.6 & $-0.108 \pm 0.028$ & $-0.41 \pm 0.67$ \\
NGC 6558 Field 3 & 6.0 & $-0.126 \pm 0.025$ & $+0.02 \pm 0.60$ \\
NGC 6642 Field 3 & 6.4 & $-0.070 \pm 0.038$ & $-1.31 \pm 0.91$ \\
\tableline
\end{tabular}
\end{center}
\caption{BULGE FIELD GIANT BRANCH SLOPES AND METALLICITIES}
\end{table*}

\clearpage

\begin{center}
FIGURE CAPTIONS
\end{center}

\figcaption
[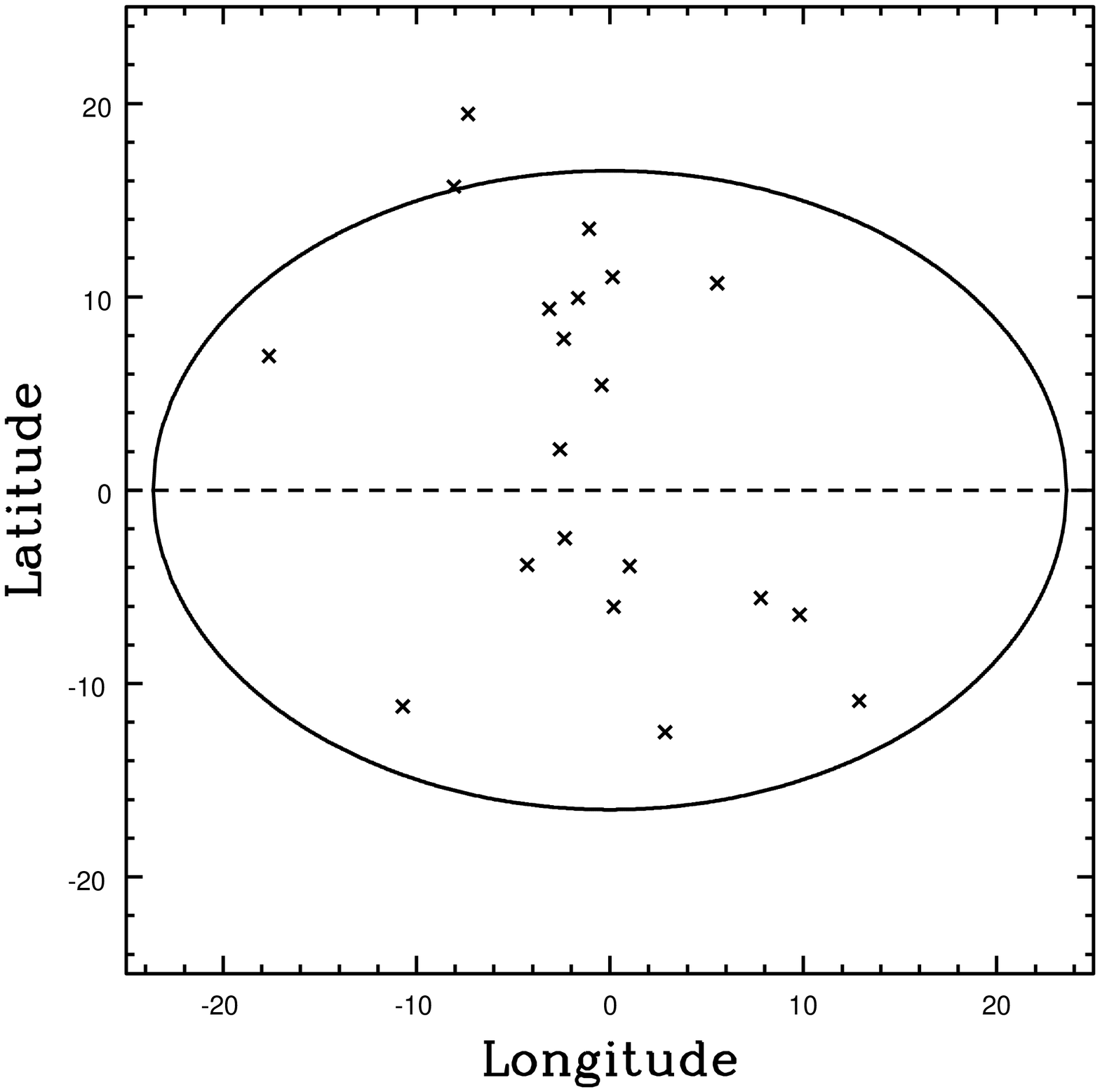]
{The projected distribution of metal-poor inner spheroid clusters. 
With the exceptions of Djorgovski 1 and HP 1, the 
clusters were selected to have R$_{GC} \leq 3.5$ kpc and [Fe/H] $\leq -1.3$ 
according to the 1996 version of the Harris (1996) database. The dashed line 
shows the Galactic Plane, while the solid line shows an ellipse with axial 
ratio 0.7, which is the ellipticity of the bulge measured by Blanco \& 
Terndrup (1989). Note that the projected distribution of the cluster sample 
differs from that of the bulge, likely due to incompleteness 
near the Galactic Plane.}

\figcaption
[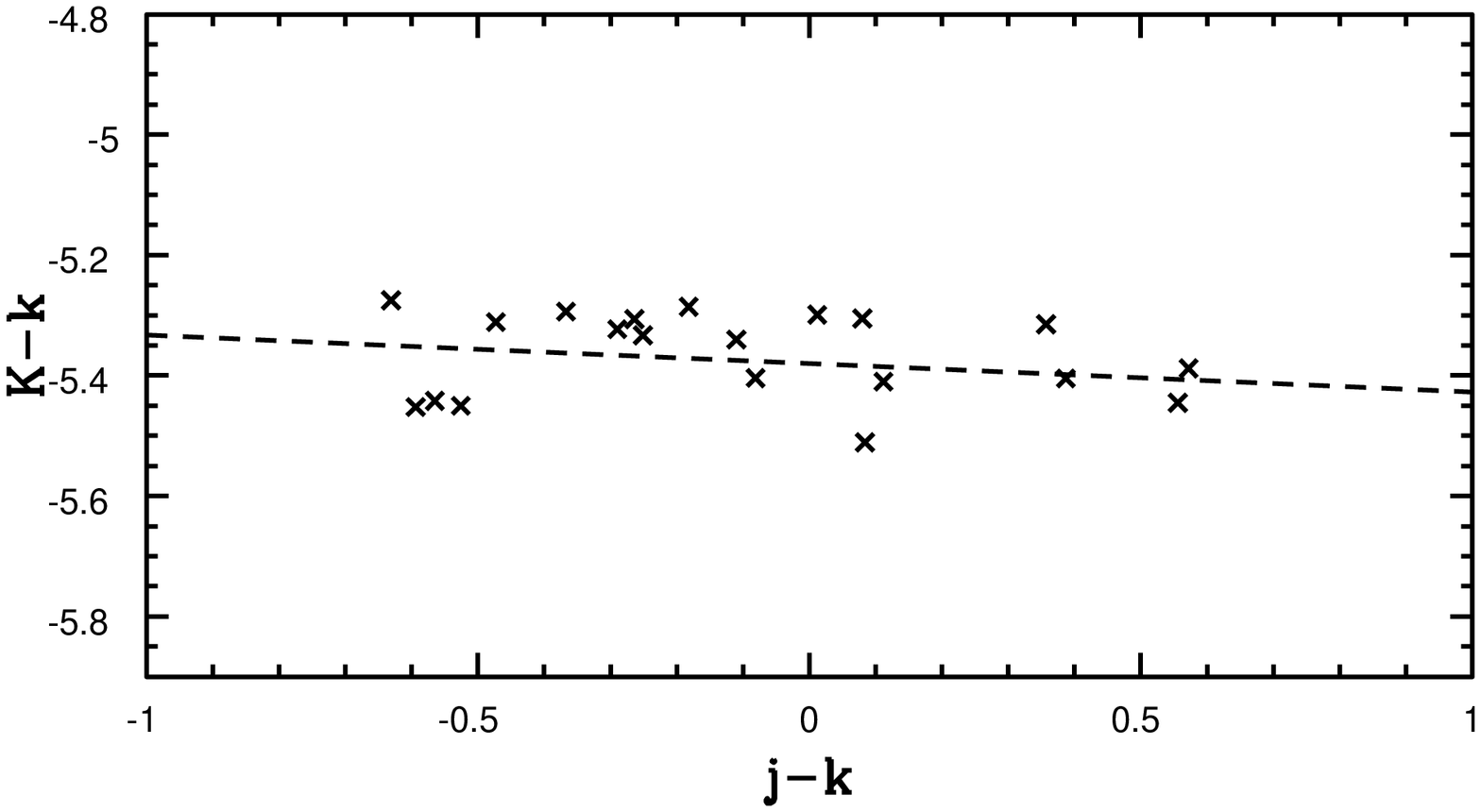]
{The difference between the standard and instrumental brightness in $K$ 
as a function of instrumental color for the Casali \& Hawarden (1992) standard 
stars. The dashed line shows a least squares fit through the datapoints.}

\figcaption
[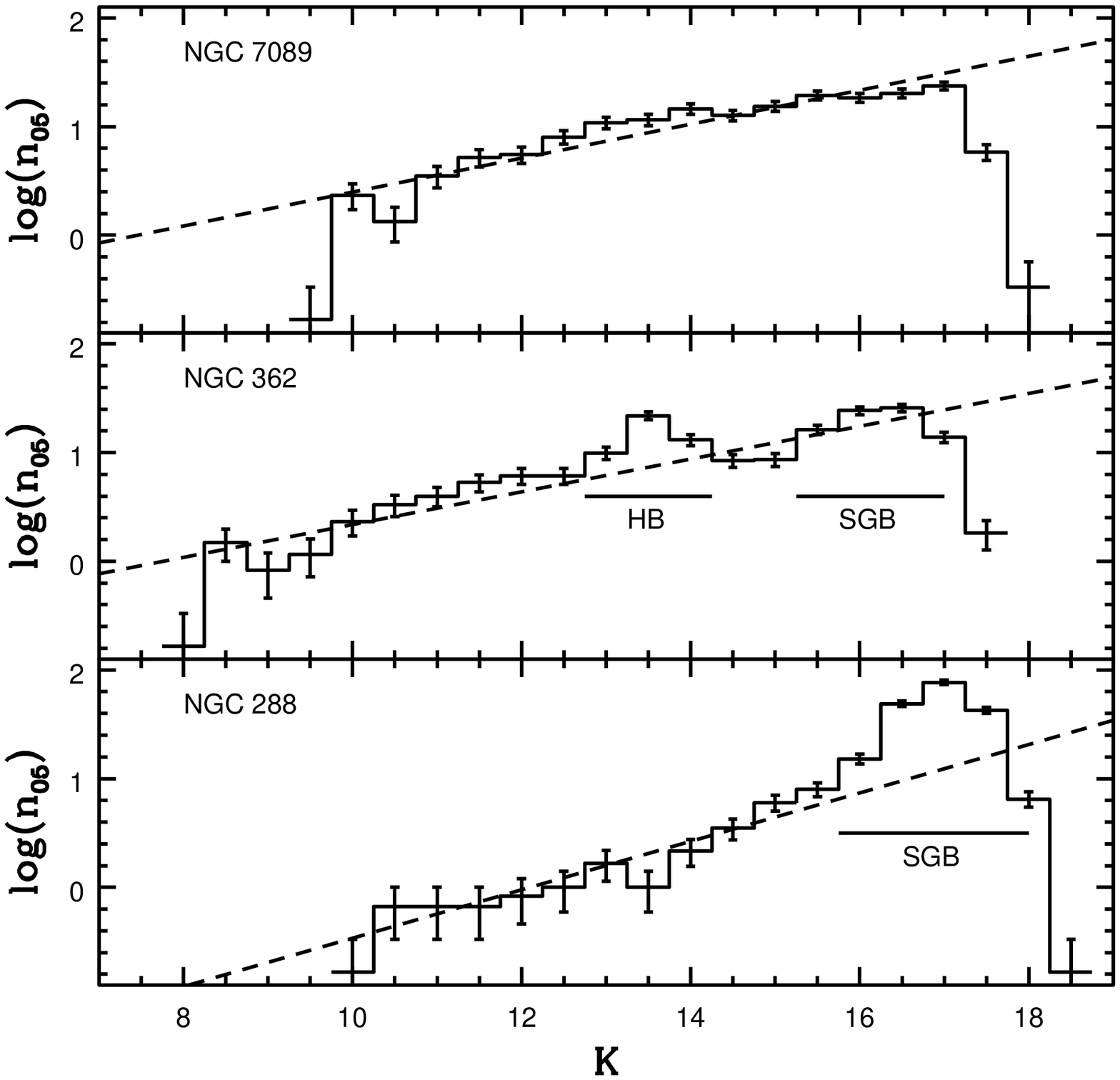]
{The $K$ LFs of NGC 7089, NGC 362, and NGC 288. n$_{05}$ is the 
number of stars per square arcmin per 0.5 mag interval. The dashed lines show 
power-law fits to the LFs, with the exponents listed in Table 2. The intervals 
corresponding to the HB and SGB in NGC 362, and the SGB in NGC 288, which were 
excluded from the fits, are indicated.}

\figcaption
[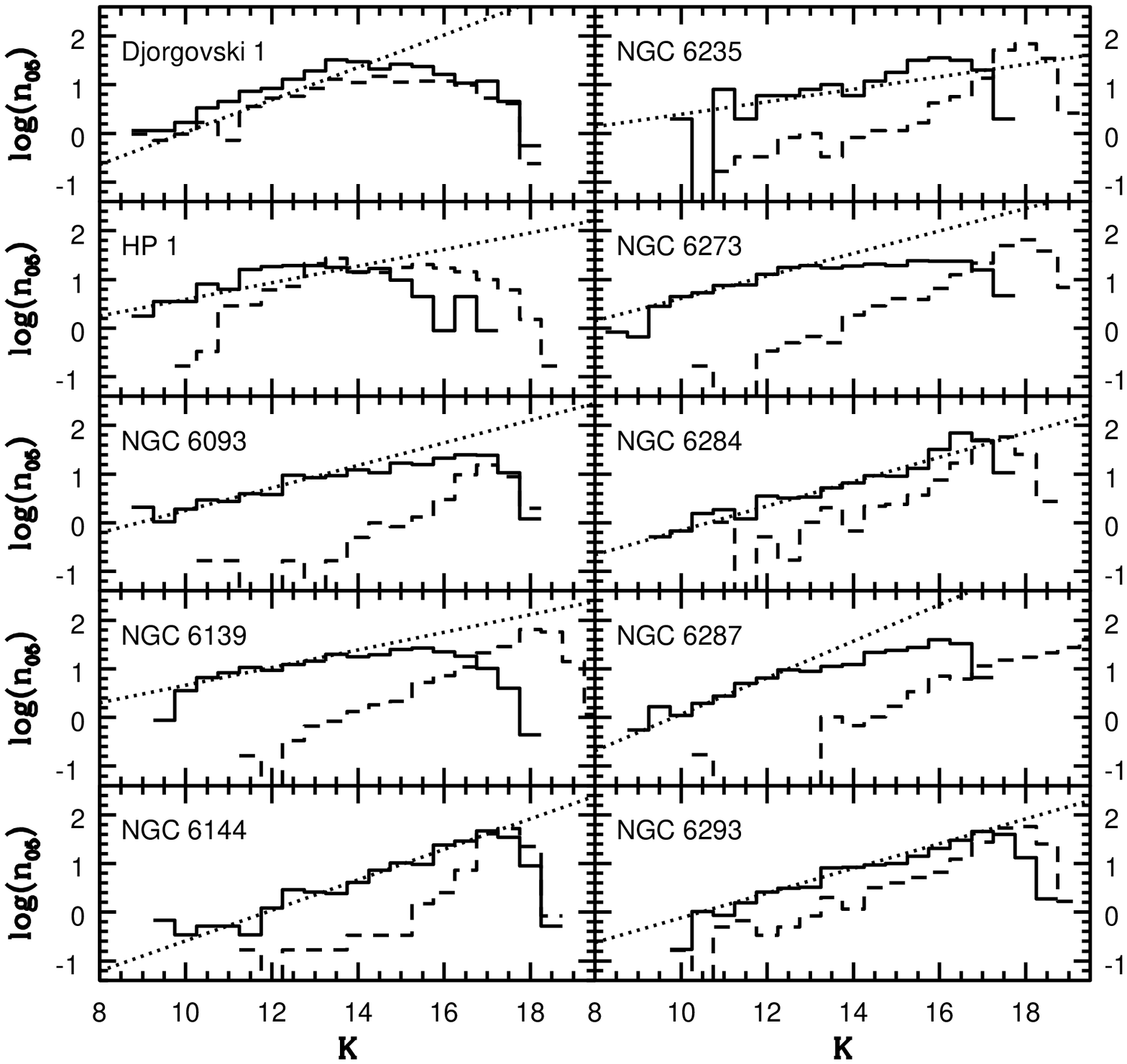,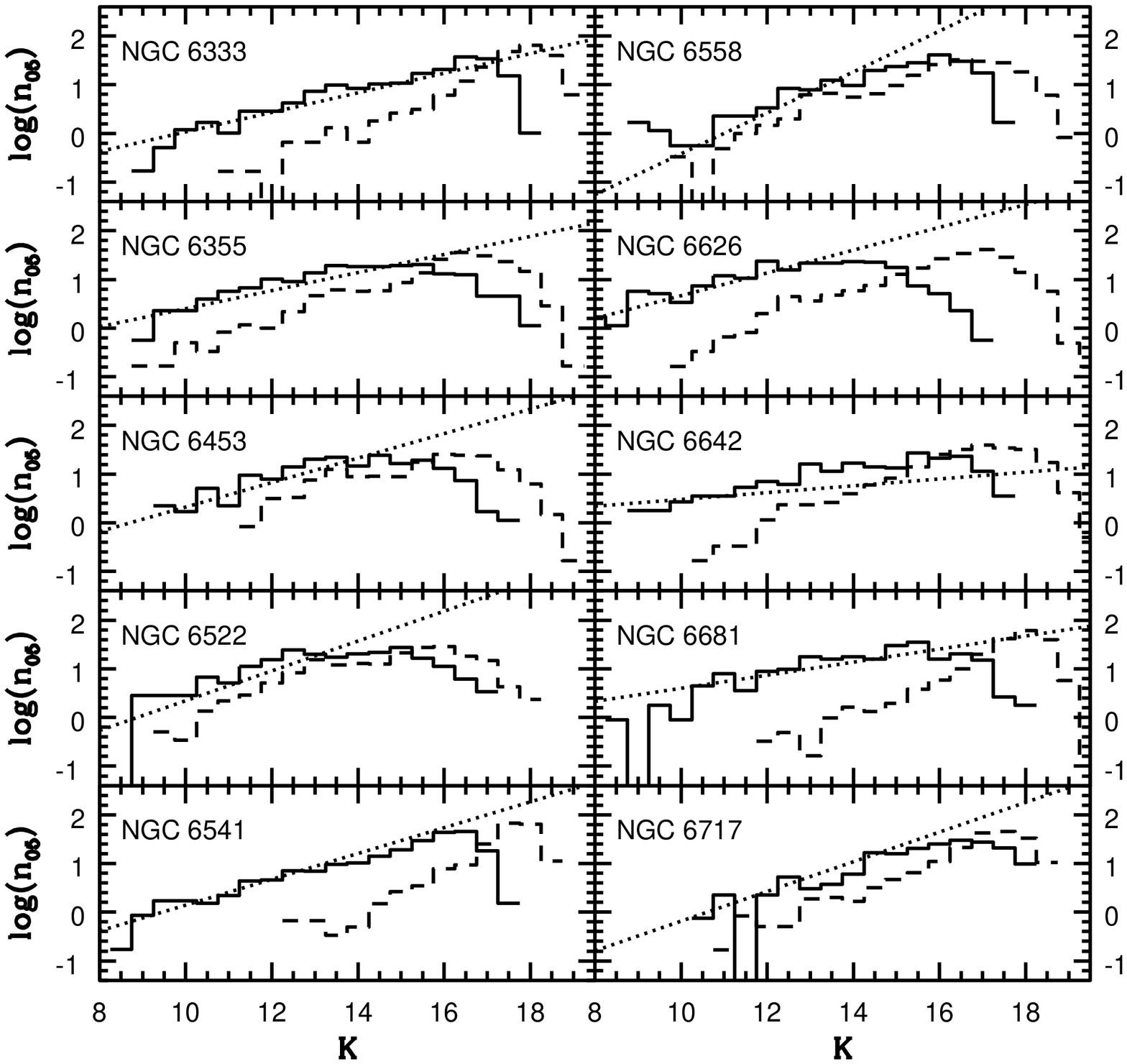]
{The inner cluster field (solid lines) and Field 3 (dashed lines) $K$ LFs. The 
dashed lines shown for Djorgovski 1 and NGC 6717 are the LFs of the outer 
cluster field and Field 2, respectively. n$_{05}$ is 
the number of stars per square arcmin per 0.5 mag interval. Stars 
brighter than $K \sim 10$ are saturated in the bulge field datasets. 
The dotted lines show least squares fits made to the inner cluster field data 
in the interval $10 \leq K \leq 12.5$ after subtracting the Field 3 
LFs, and the power-law exponents for these fits are listed in Table 3.}

\figcaption
[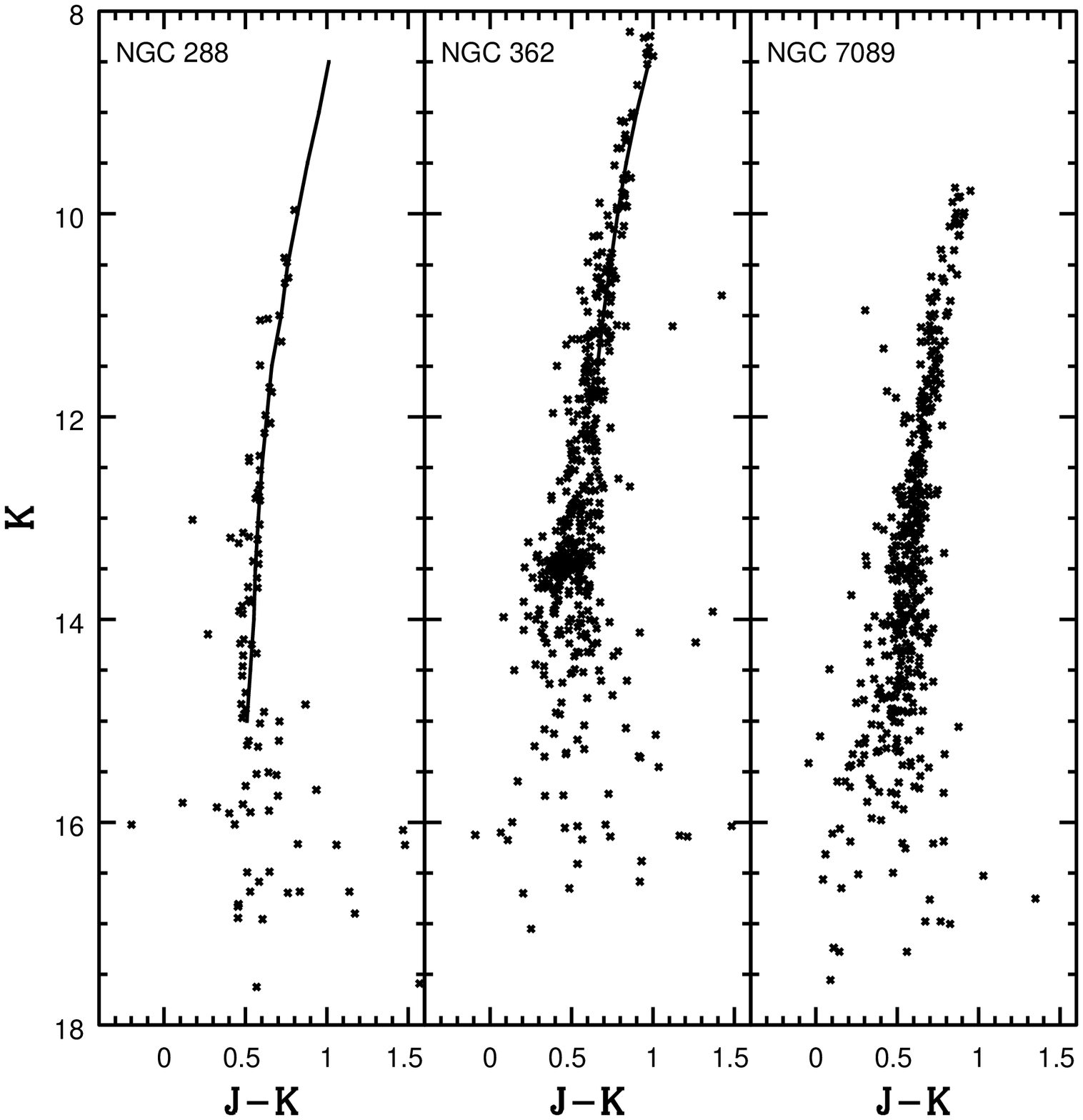]
{The $(K, J-K)$ CMDs of NGC 288, NGC 362, and NGC 7089. The lines 
show the loci of data published by Davidge \& Harris (1997) for NGC 288, and 
Frogel et al. (1983) for NGC 362.}

\figcaption
[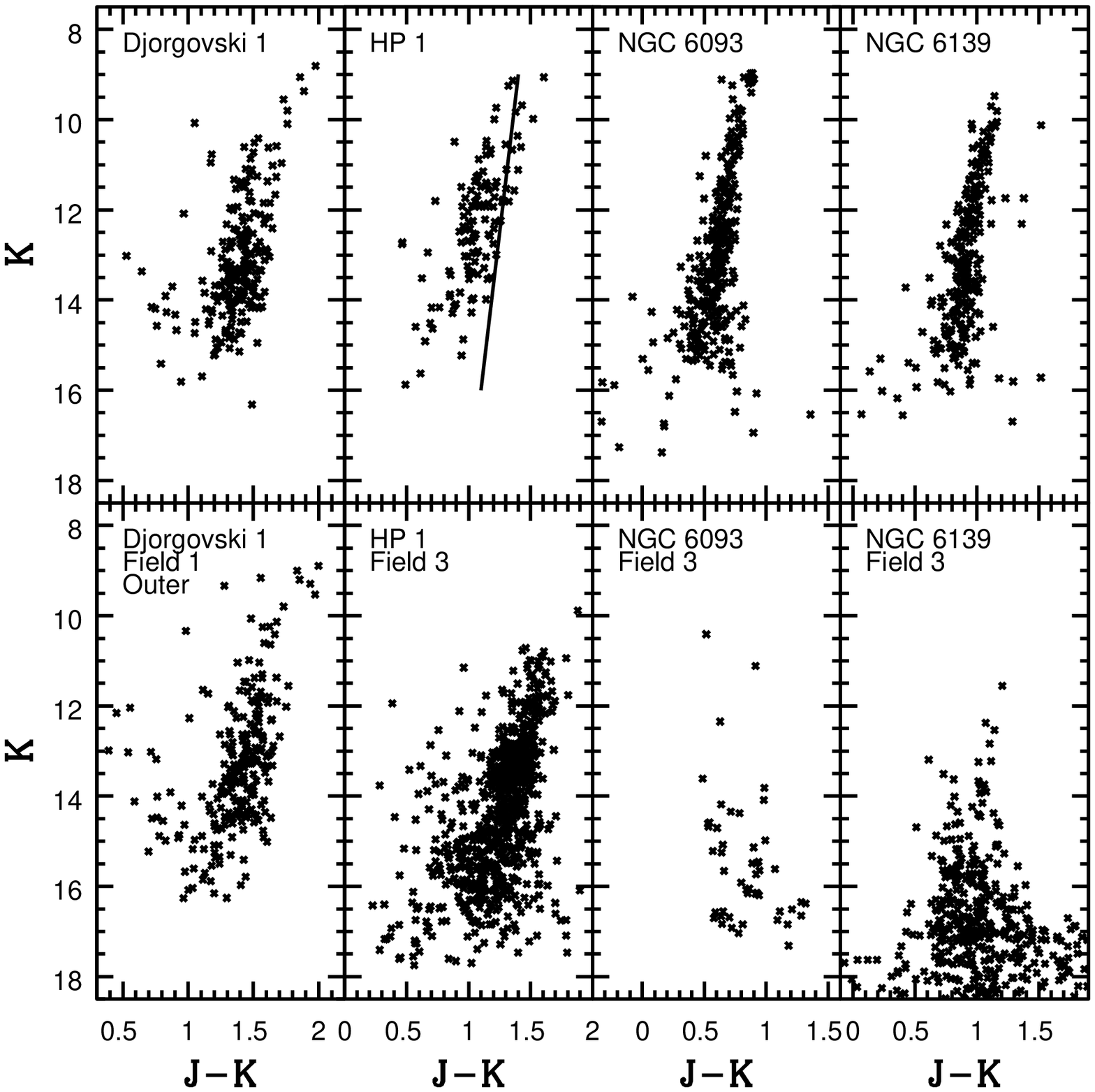,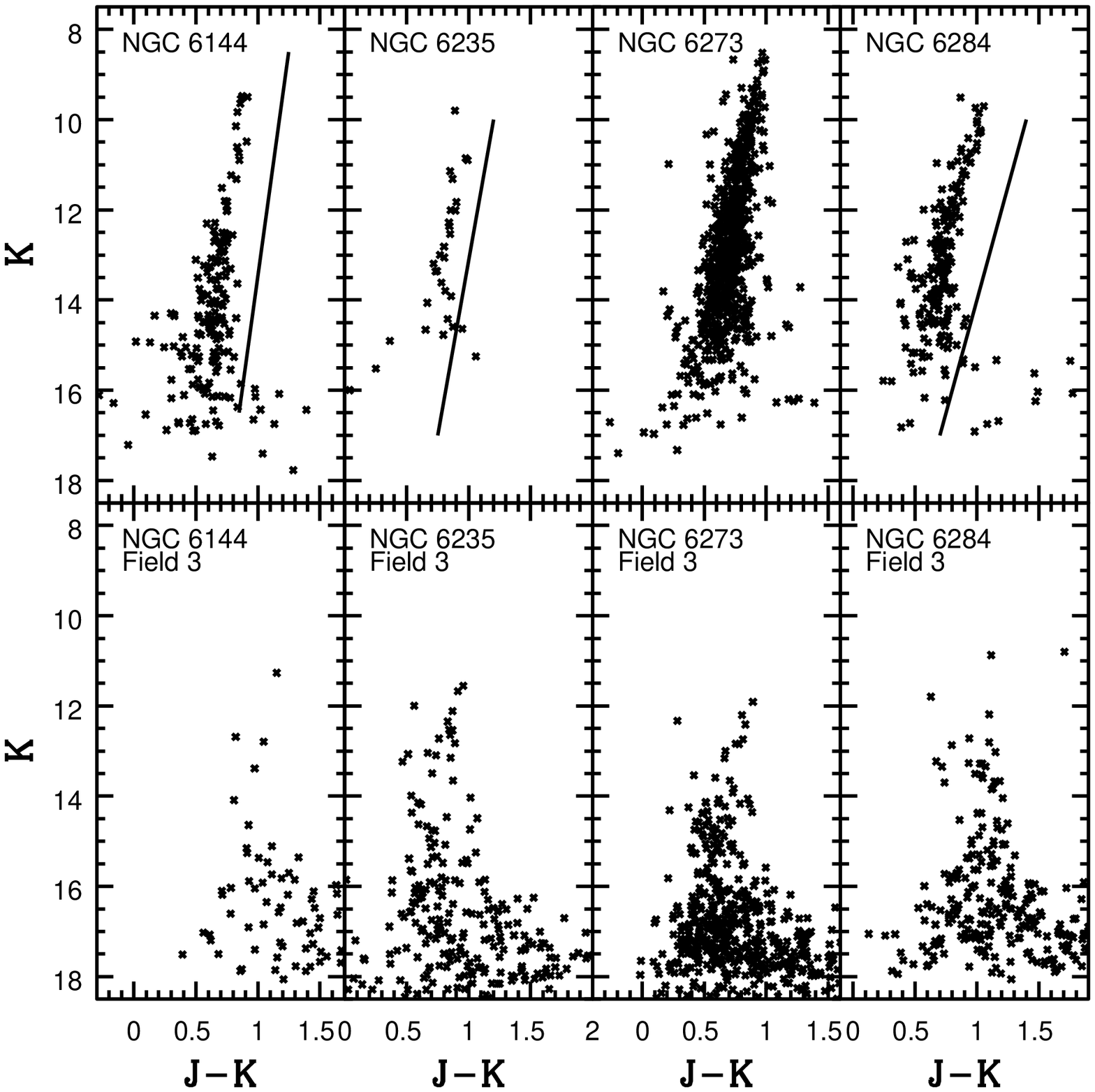,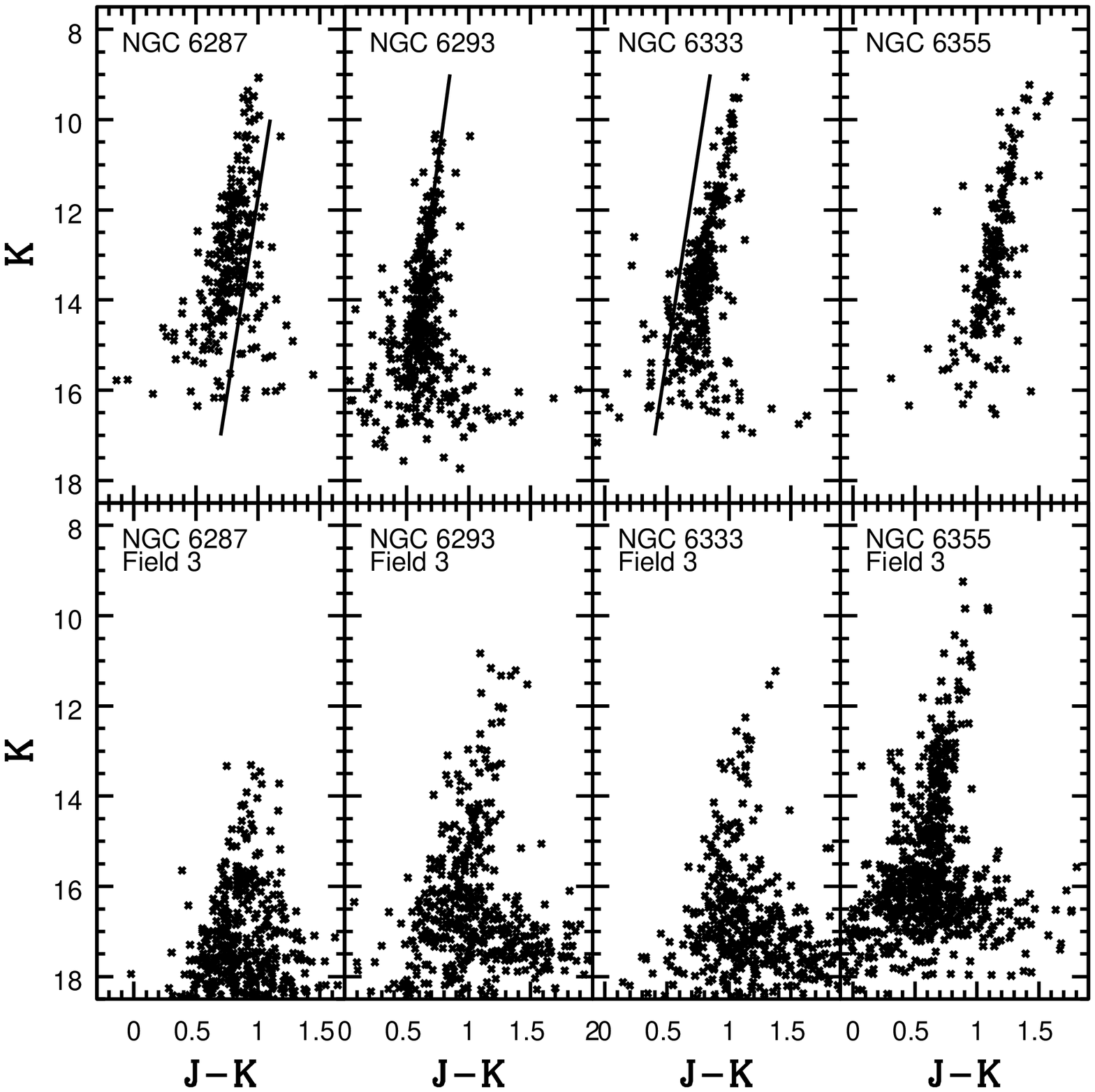,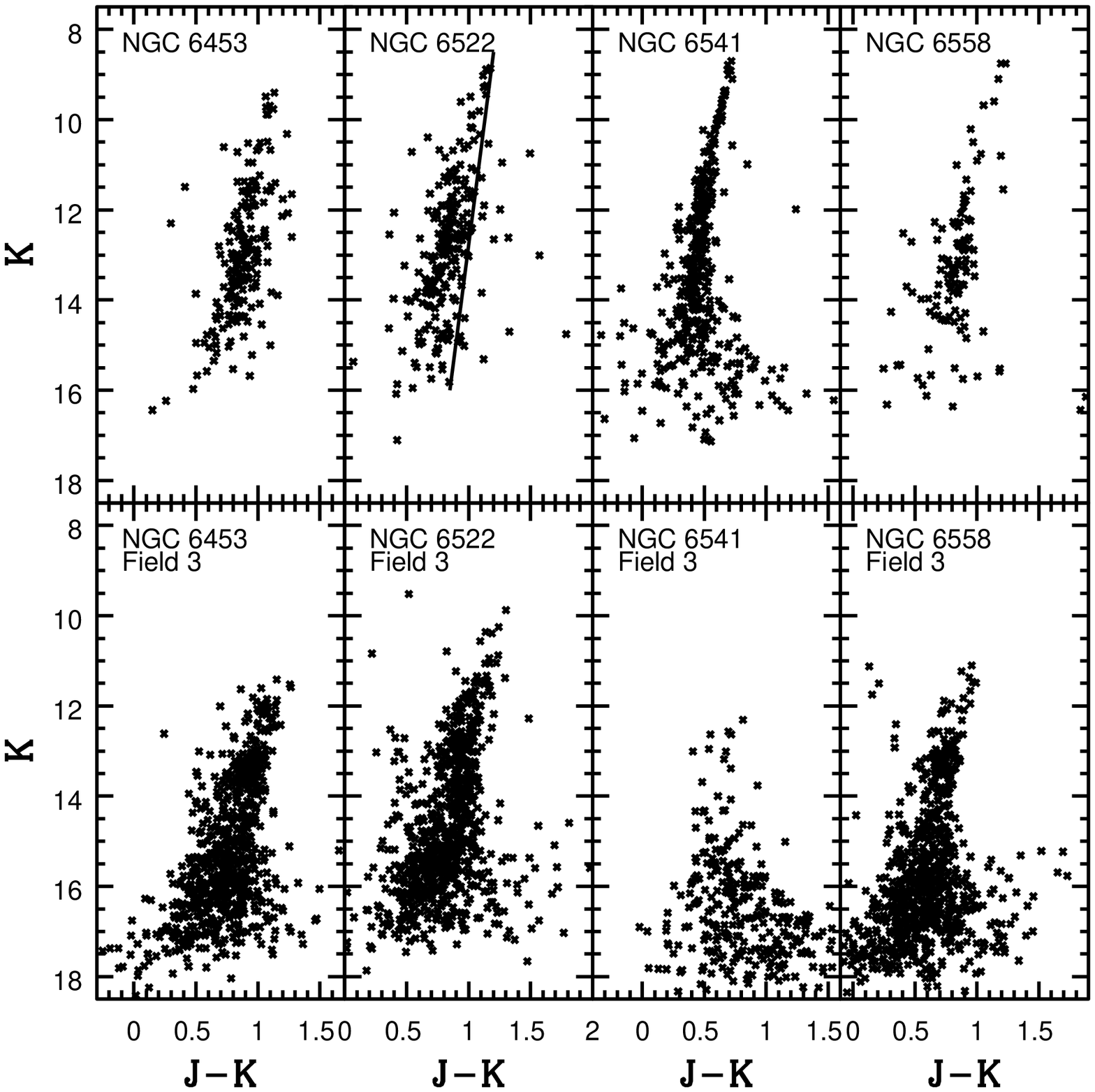,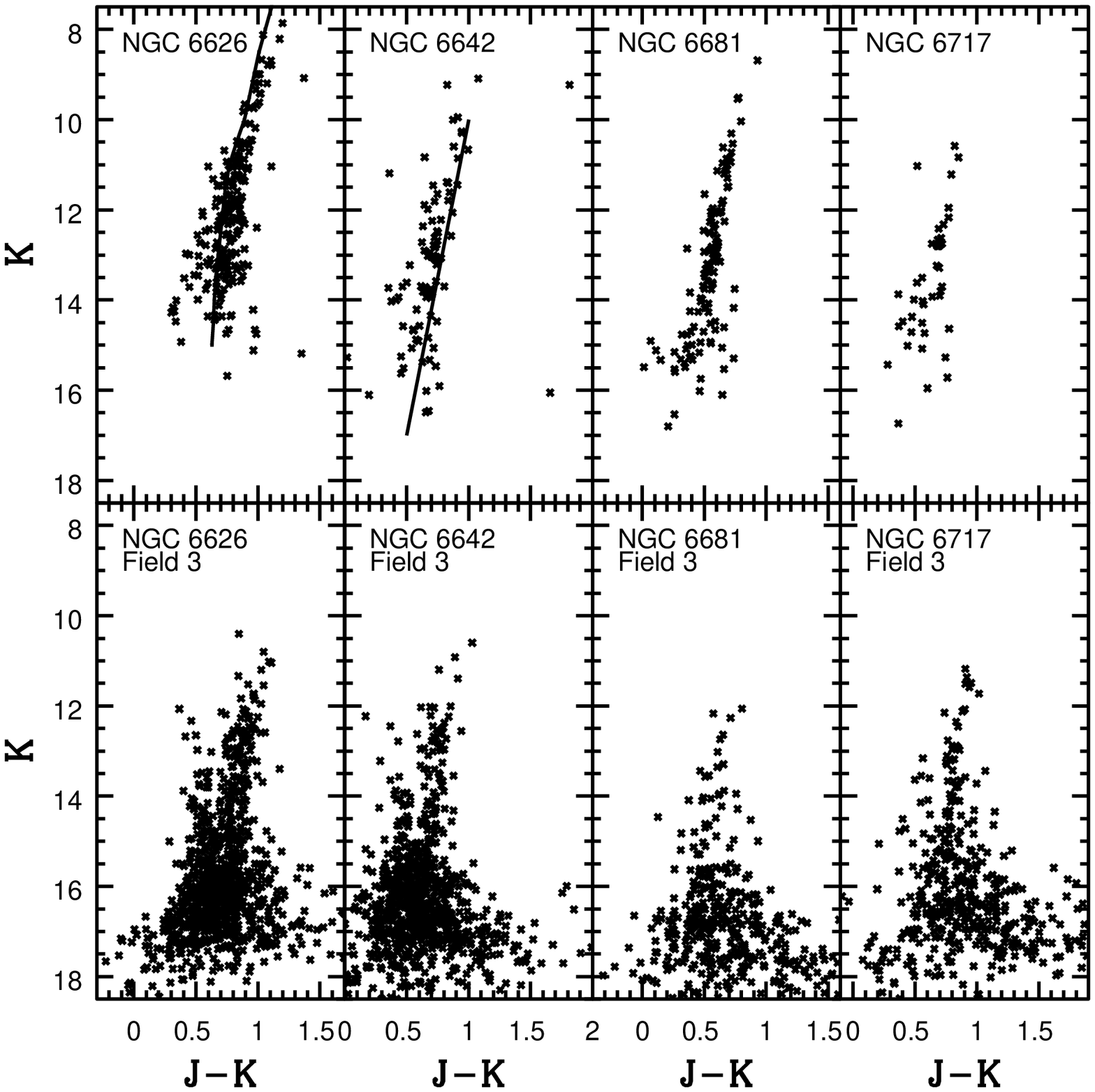]
{The $(K, J-K)$ CMDs of the inner spheroid clusters and 
bulge fields. The solid lines superimposed on the HP 1, NGC 6144, NGC 
6235, NGC 6284, NGC 6287, NGC 6293, NGC 6333, NGC 6522, and NGC 6642 CMDs are 
the trends defined by Minniti et al. (1995a). The solid 
line in the NGC 6626 panel shows the normal points from Table 3 of 
Davidge et al. (1996).}

\figcaption
[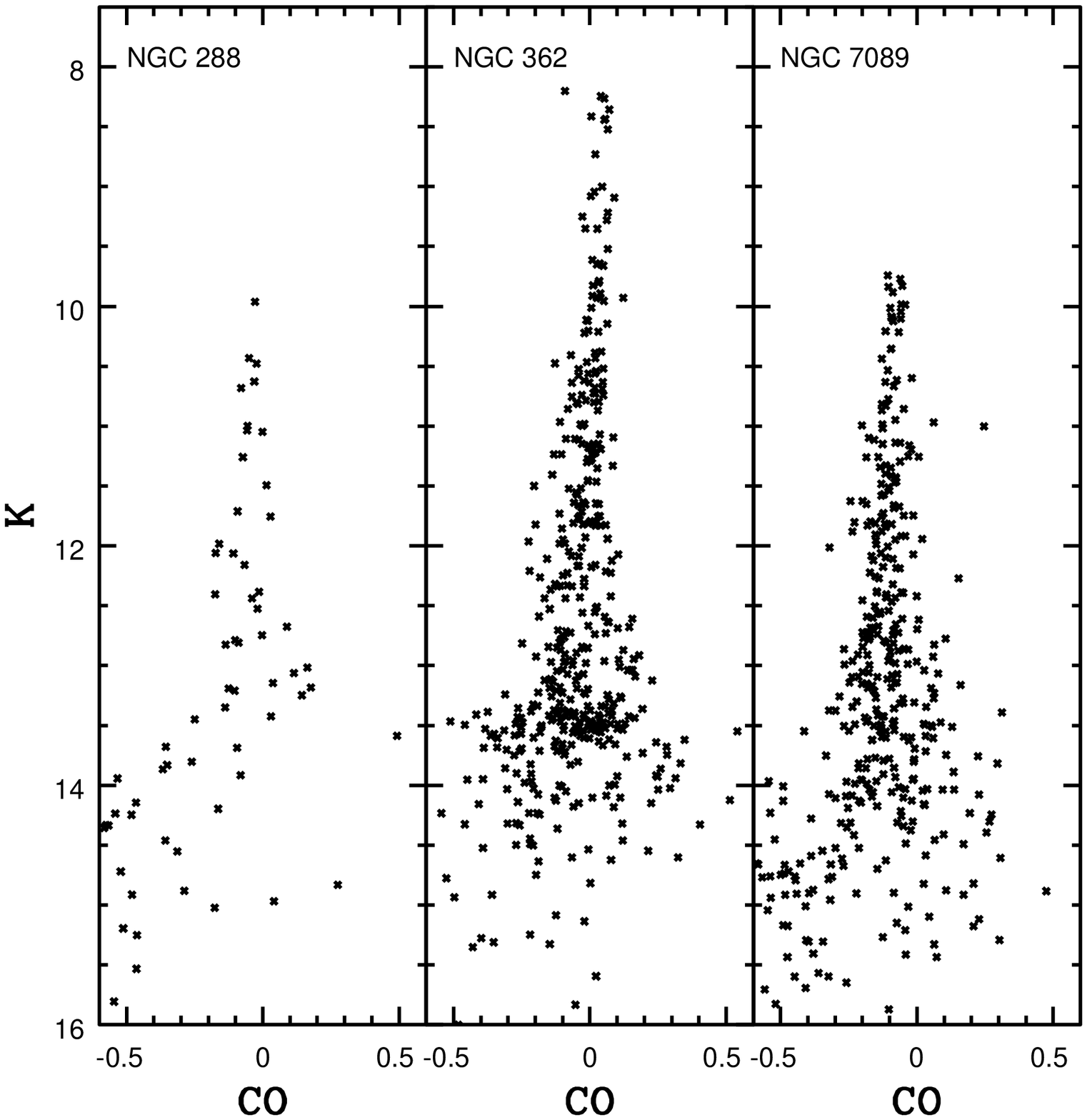]
{The $(K, CO)$ CMDs of NGC 288, NGC 362, and NGC 7089. The stars 
plotted in this figure have been detected in all 5 filters.}

\figcaption
[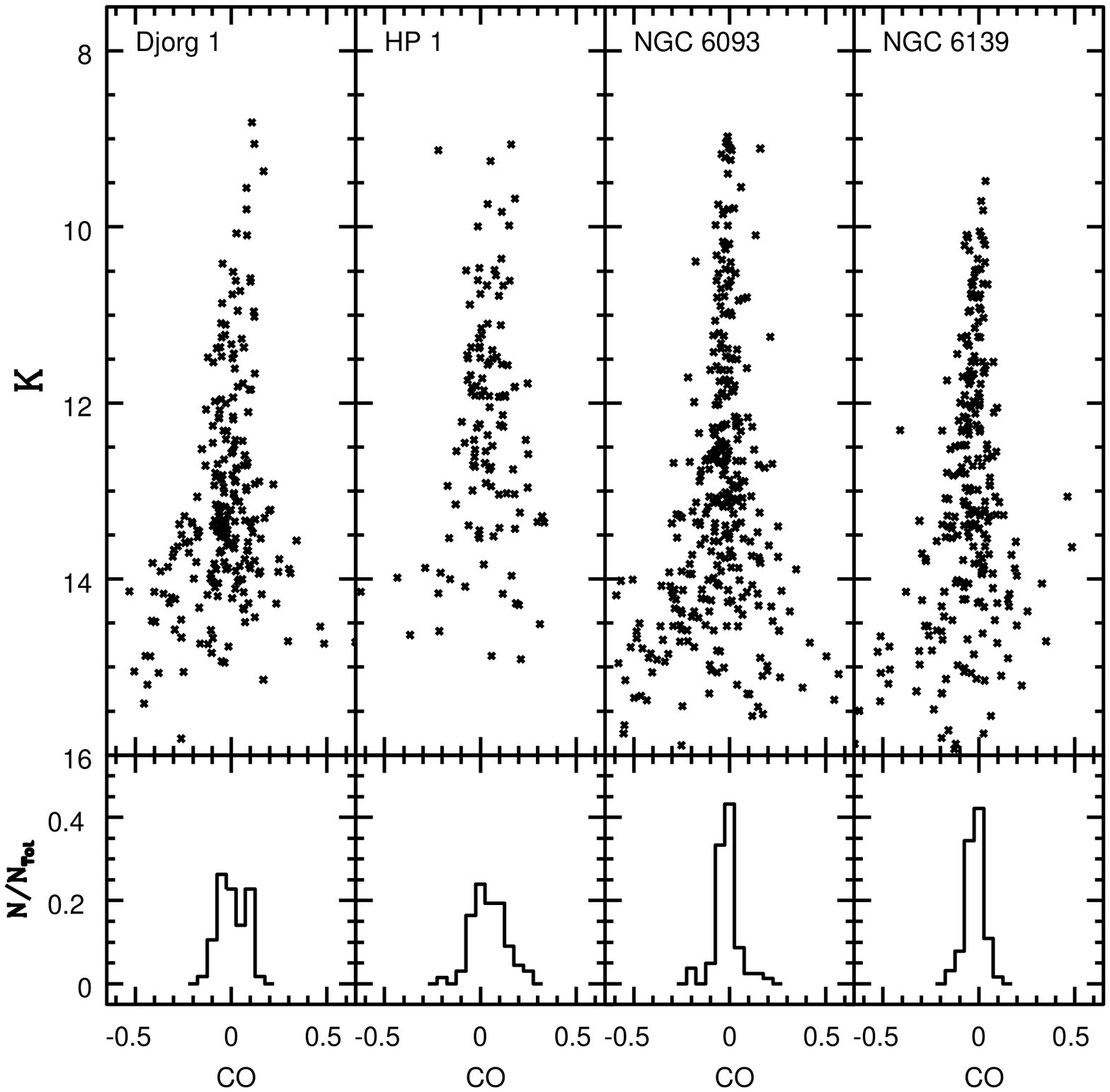,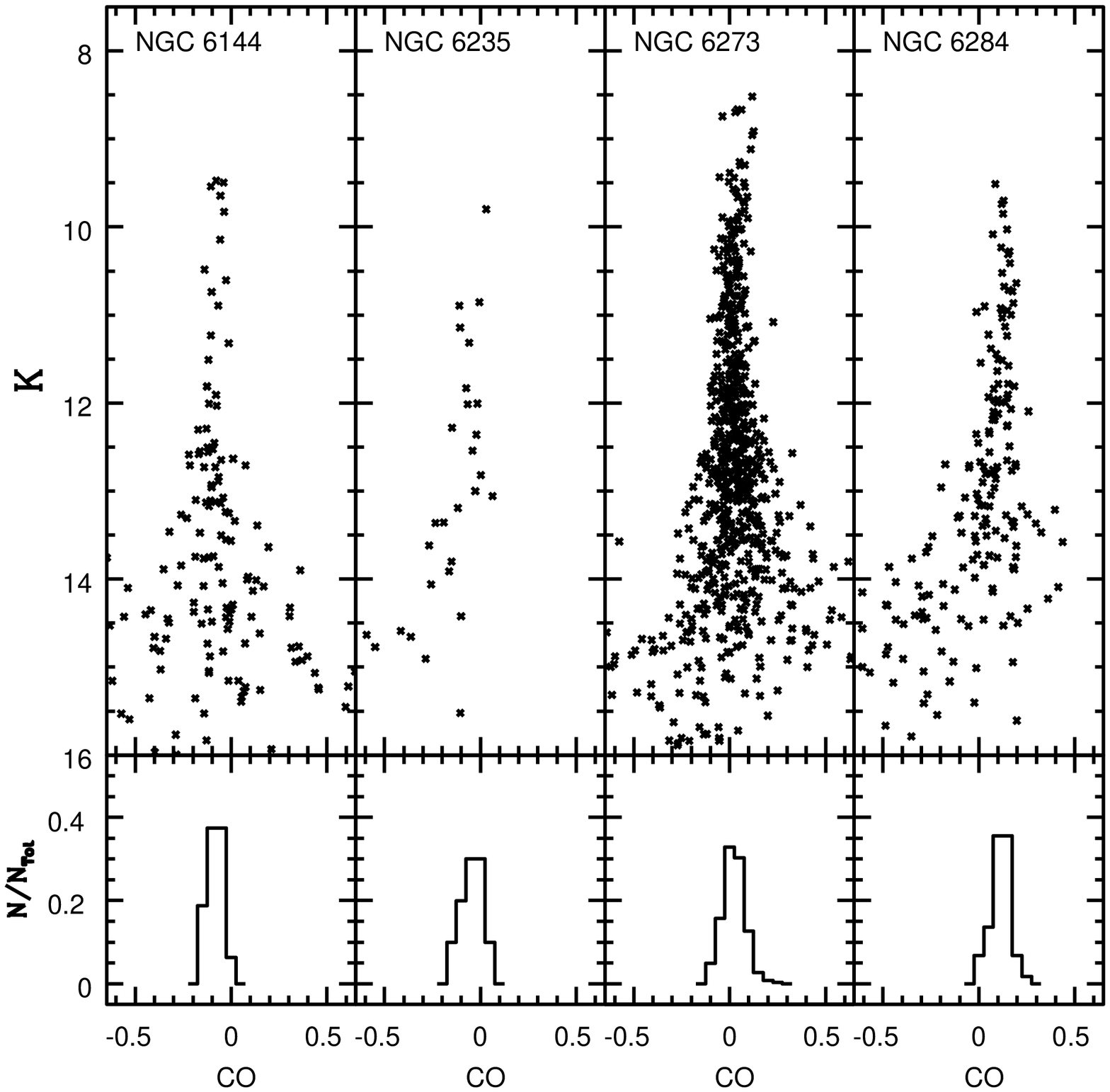,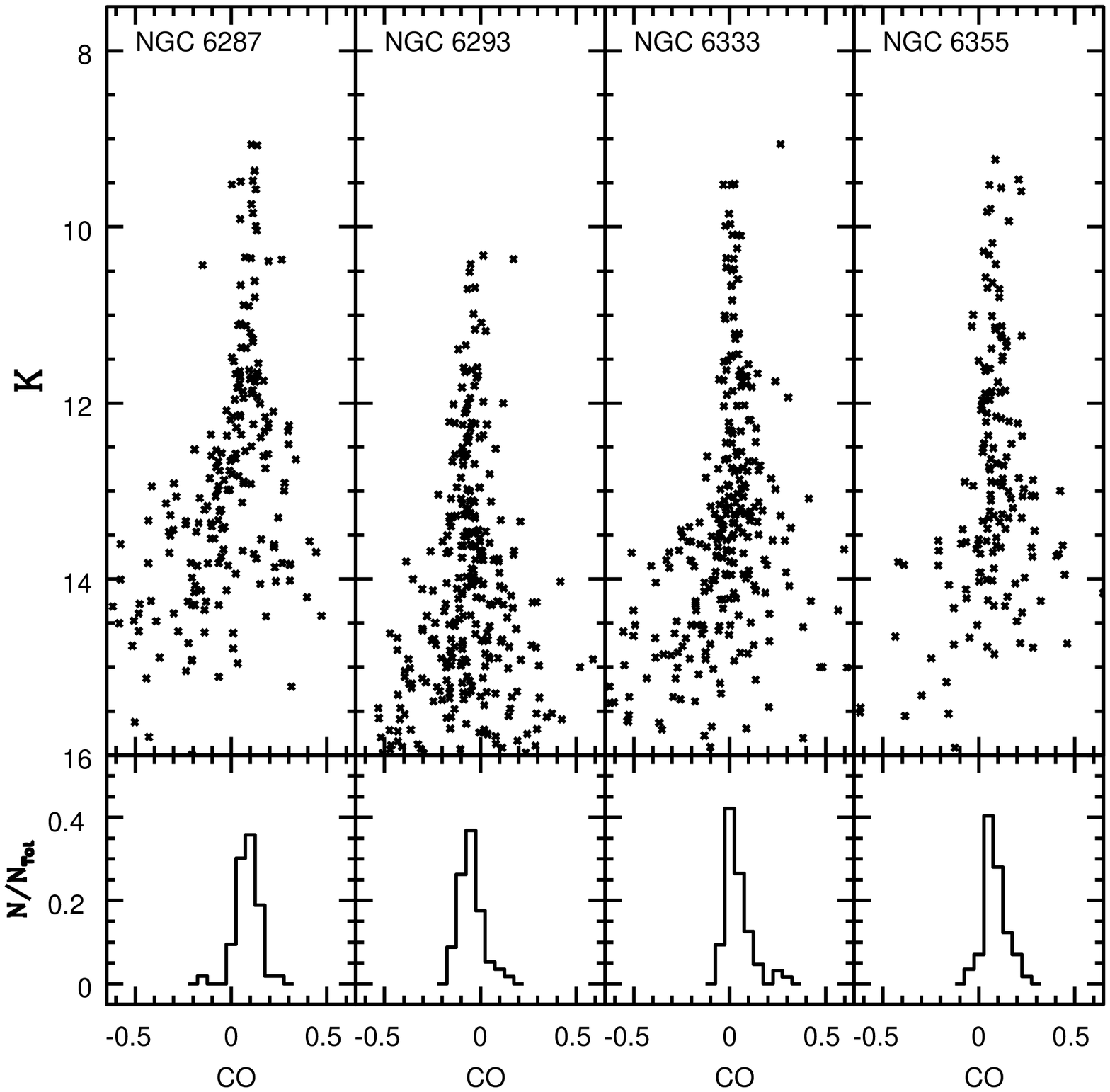,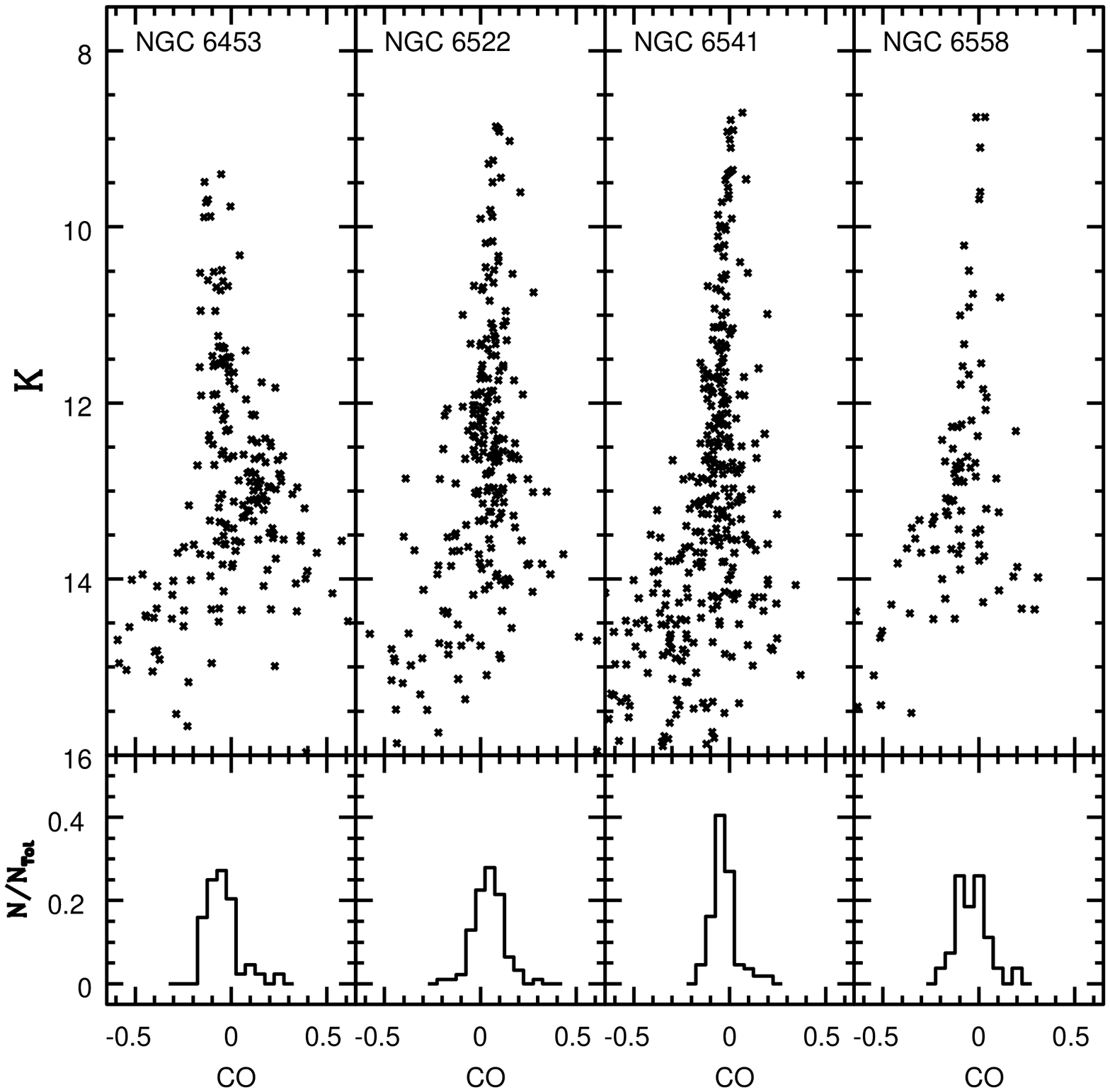,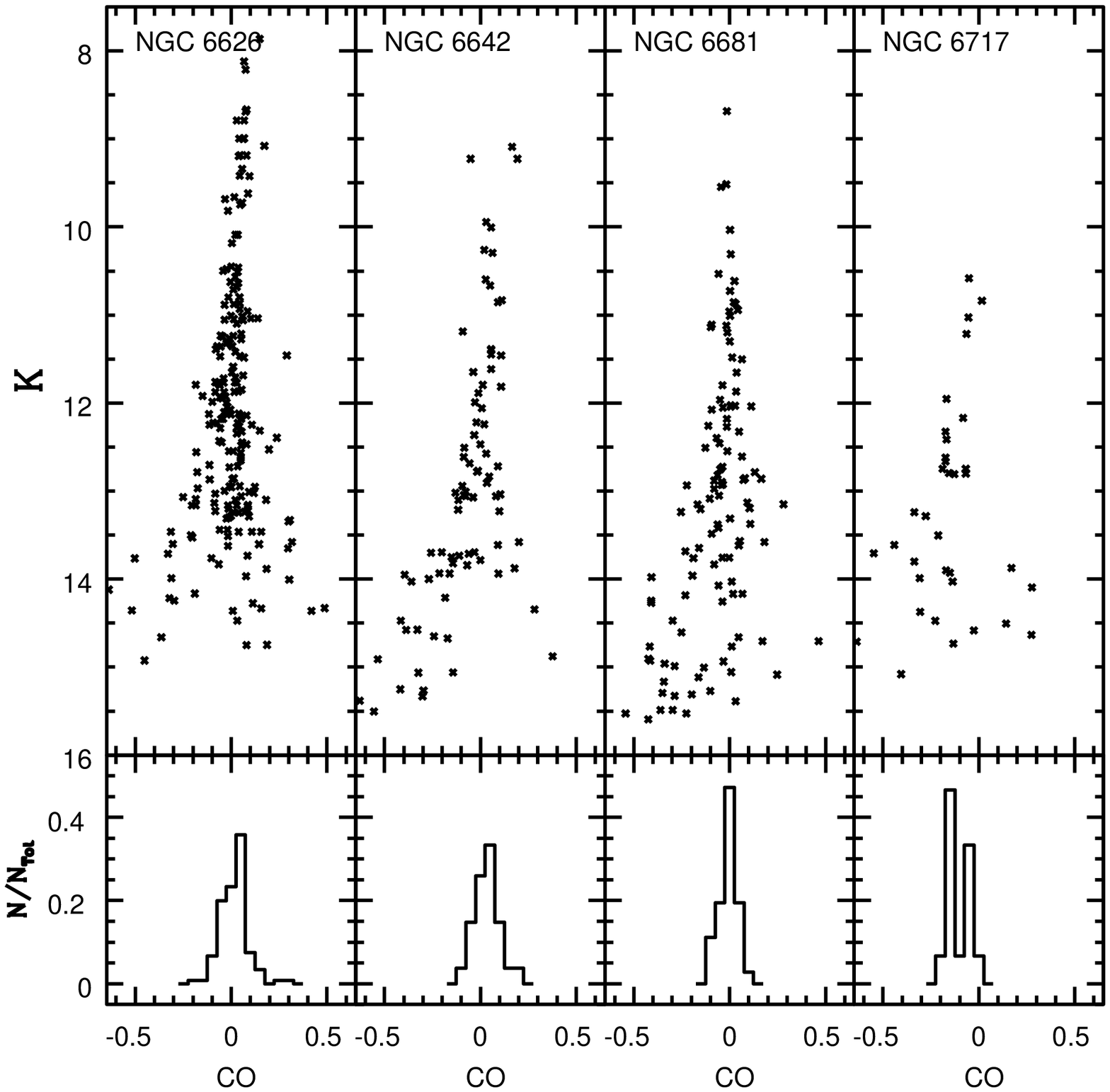]
{The $(K, CO)$ CMDs of the inner cluster fields (top panels) and the 
histogram distributions of CO indices for stars brighter than $K = 12.5$ (lower 
panels), normalized according to the total number of stars with $K \leq 12.5$ 
(N$_{Tot}$). Note that the CO distributions for the clusters with 
the lowest $C$ values tend to be broad and, in some cases, bimodal, due to 
contamination from bulge stars.}

\figcaption
[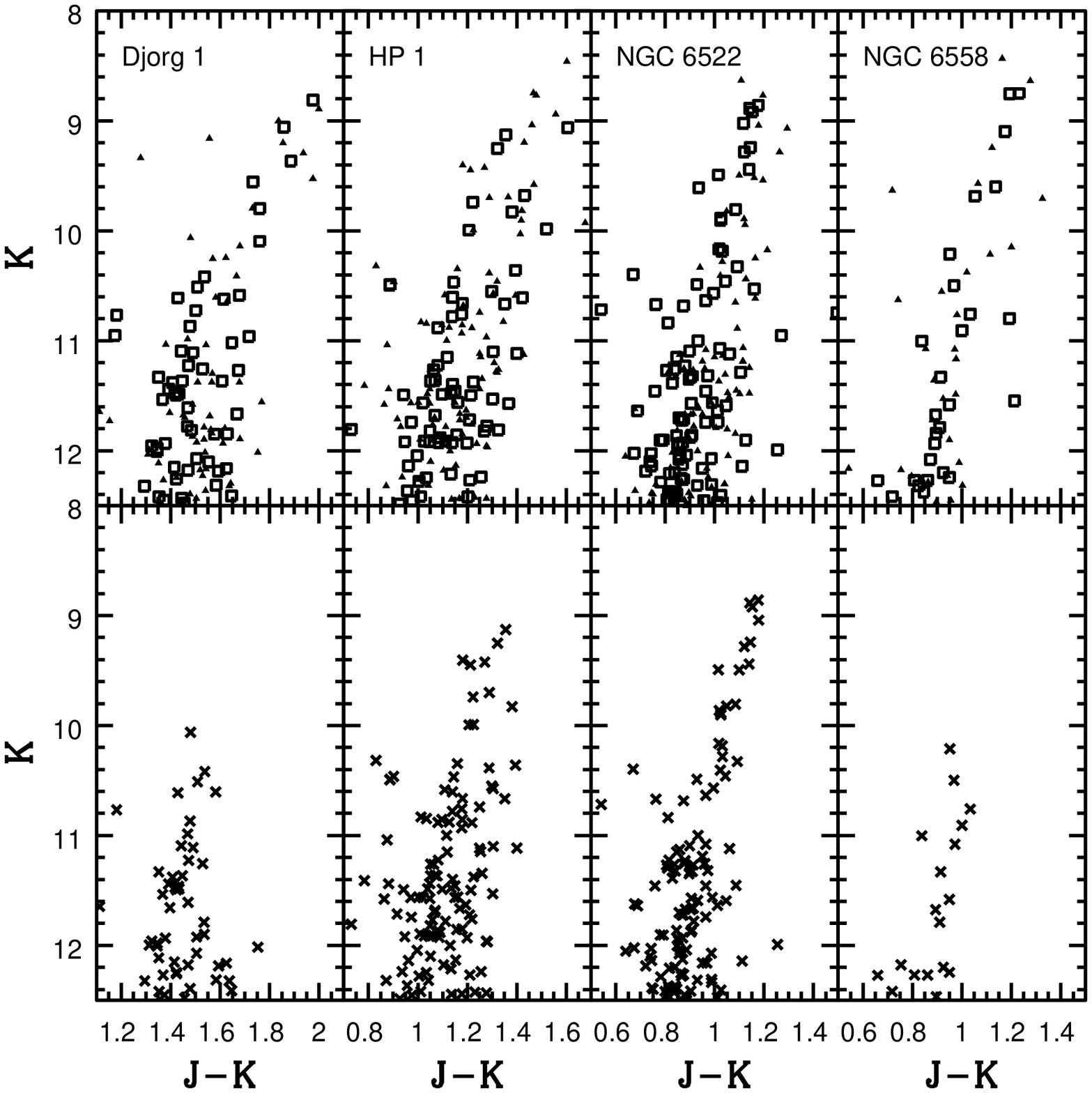]
{The $(K, J-K)$ CMDs of stars in the inner and outer cluster 
fields of Djorgovski 1, HP 1, NGC 6522, and NGC 6558, which are the clusters 
in which large numbers of bulge stars were identified. 
The top panel shows the CMDs before removing bulge stars; 
stars in the inner and outer fields are plotted as open squares and filled 
triangles, respectively. The space density of bright stars 
in the inner and outer fields are comparable, indicating that the 
majority of these objects do not follow the cluster light profiles, as expected 
if these objects belong to the bulge. The lower panels show the CMDs 
after the removal of bulge stars.}

\figcaption
[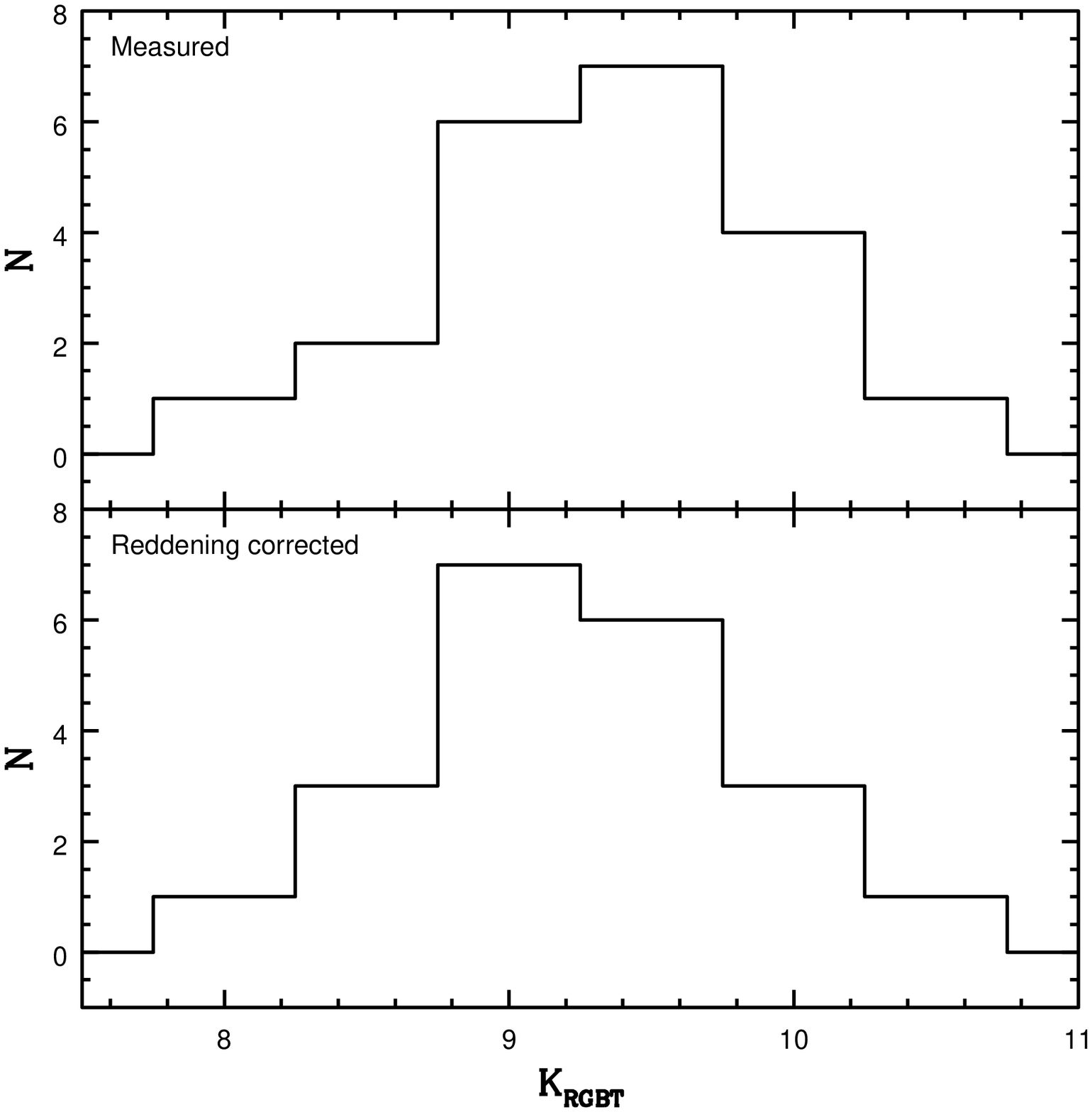]
{The histogram distribution of RGB-tip brightnesses, both 
as measured (top panel) and corrected for reddening (lower panel).} 

\figcaption
[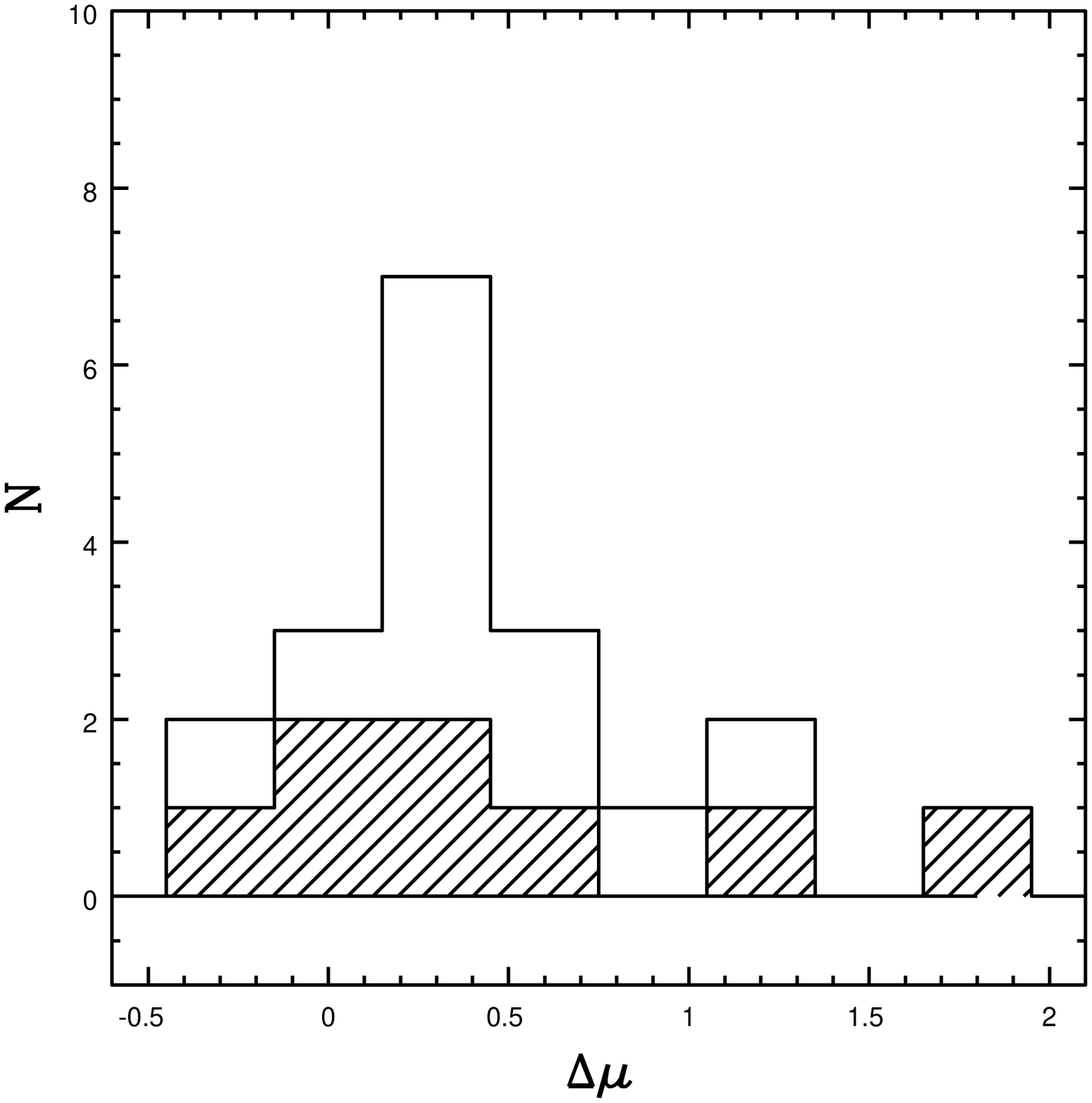]
{The histogram distribution of the difference between $\mu^{emp}_0$ and 
$\mu^{H96}_0$, which are defined in the text. The open curve shows the 
distribution for all clusters, while the shaded curve shows the distribution 
for core-collapsed systems.}

\figcaption
[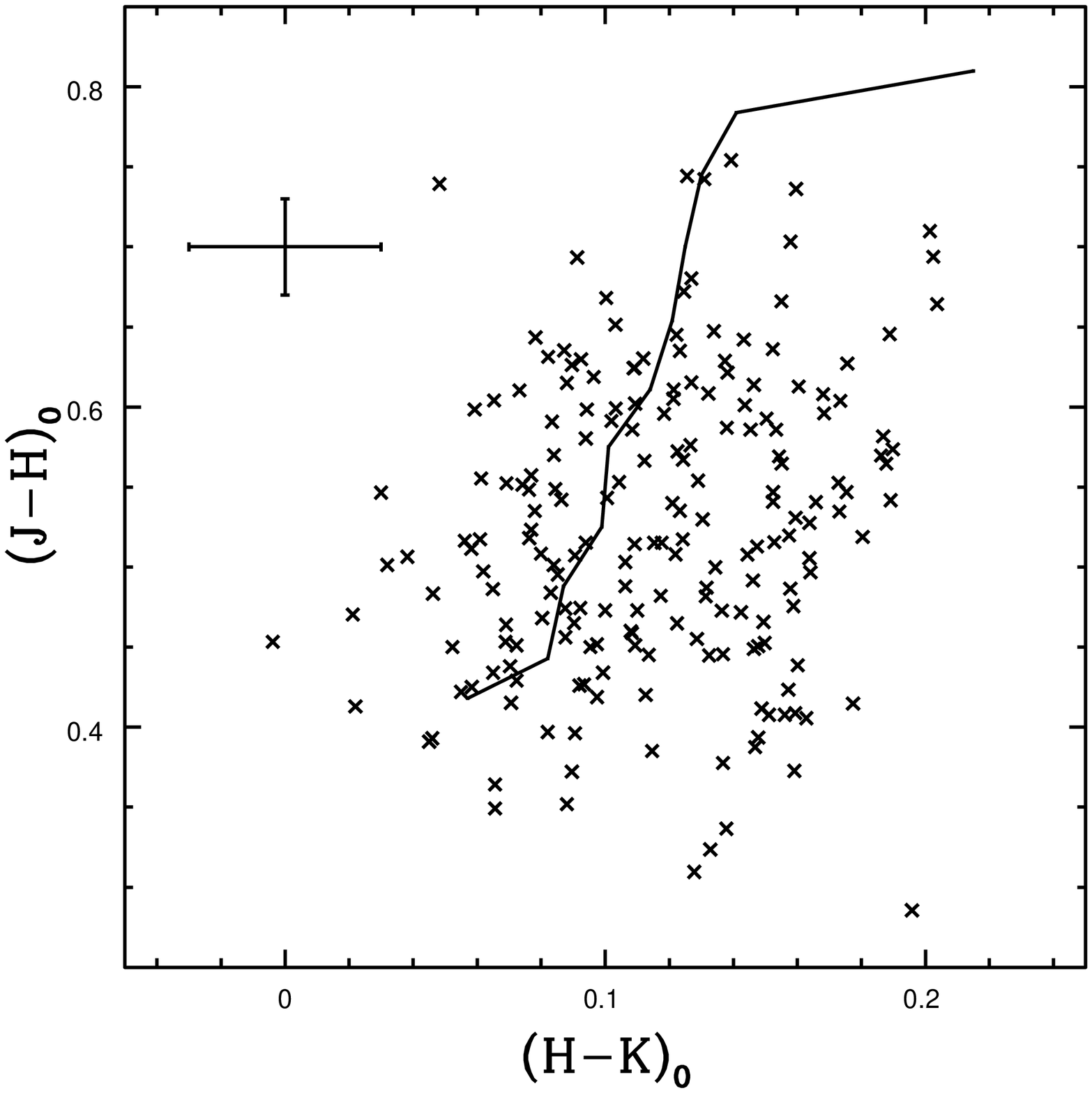]
{The $(J-H, H-K)$ TCD of the giant branches in metal-poor inner spheroid 
clusters. Normal points, which were created by computing the mode of the color 
distribution in $\pm 0.25$ mag bins along the $K$ axes of the $(K, H-K)$ and 
$(K, J-K)$ CMDs, are plotted for each cluster to reduce observational 
scatter. The points have been de-reddened using the $E(B-V)$ entries in 
column 4 of Table 6 according to the Rieke \& Lebofsky (1985) reddening 
curve. The solid line is the metal-poor halo cluster giant branch 
from Table 5 of Davidge \& Harris (1995), while the error bars show the 
uncertainties in the photometric zeropoints. Note that the 0.02 mag offset in 
$H-K$ between the midpoint of the inner spheroid data distribution and the 
halo cluster sequence falls within the uncertainties in the photometric 
zeropoints.}

\figcaption
[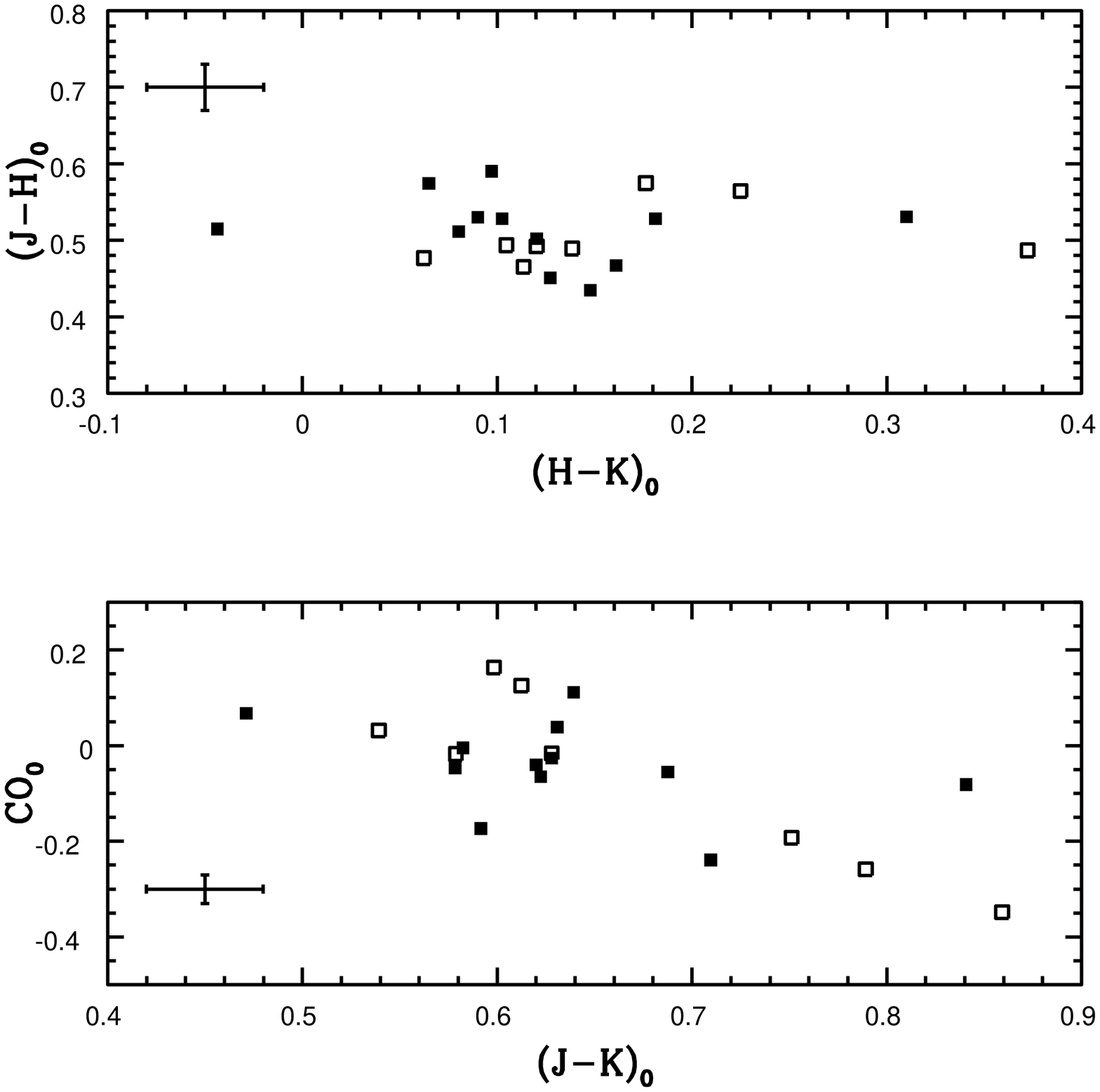]
{The $(J-H, H-K)$ and $(CO, J-K)$ TCDs for integrated light measurements. 
Clusters that Trager et al. (1995) classify as core-collapsed are 
shown as open squares, while the remaining clusters are plotted 
as filled squares. The error bars show the uncertainties in the 
photometric zeropoints.}

\figcaption
[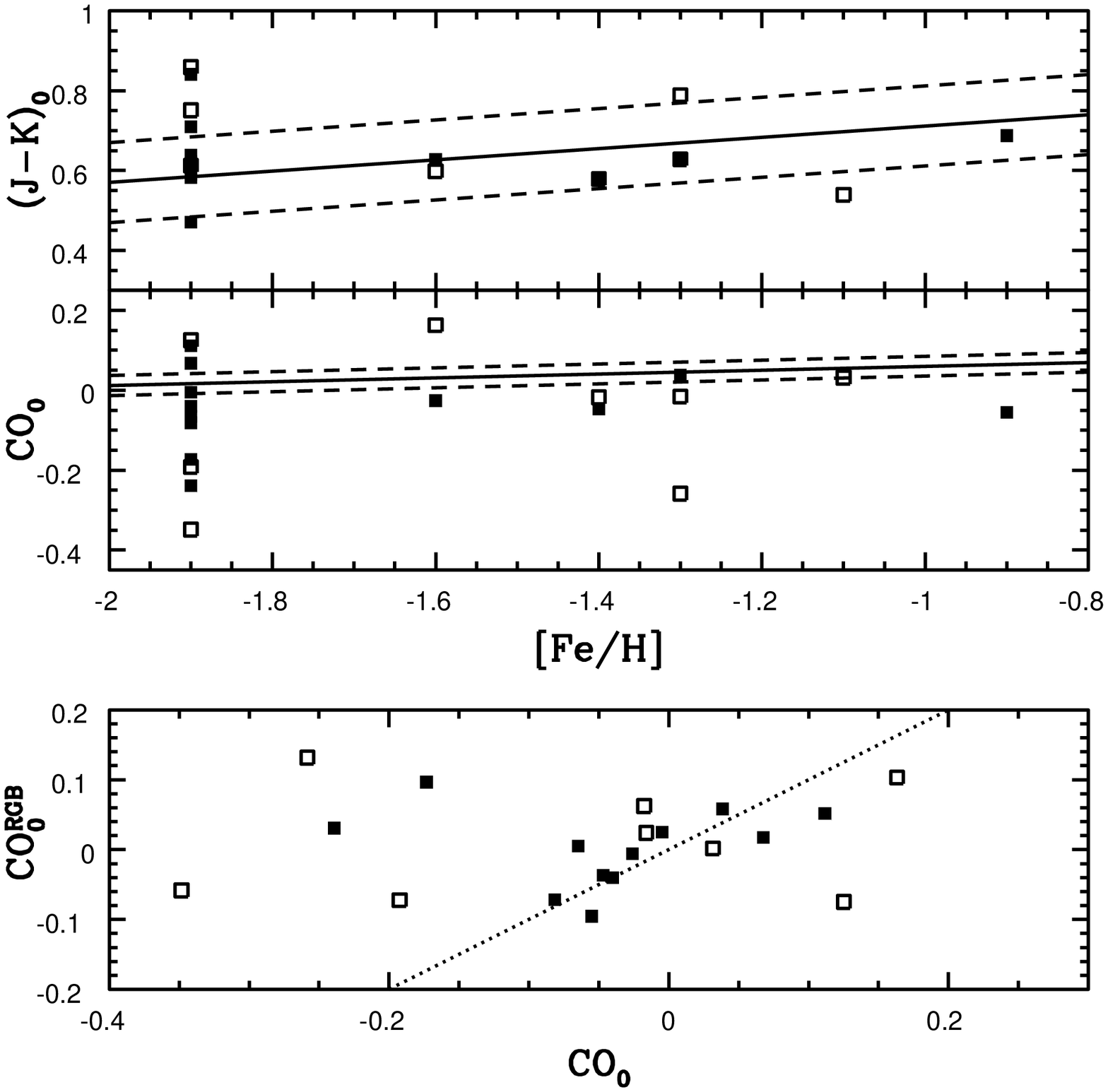]
{The top two panels examine the location of metal-poor inner 
spheroid clusters with respect to color -- metallicity 
relations defined by halo clusters. The solid lines 
show the relations defined by Aaronson et al. (1978), as re-calibrated by 
Davidge (2000); the dashed lines show the 
approximate scatter envelope in the Aaronson et al. (1978) data. 
Core-collapsed clusters are shown as open 
squares, while all other clusters are plotted as filled squares. The lower 
panel shows the CO index measured from upper giant branch stars, 
CO$^{RGB}_0$, plotted against the integrated CO index, CO$_0$. 
The dotted line shows the relation CO$^{RGB}_0$ = CO$_0$.}

\figcaption
[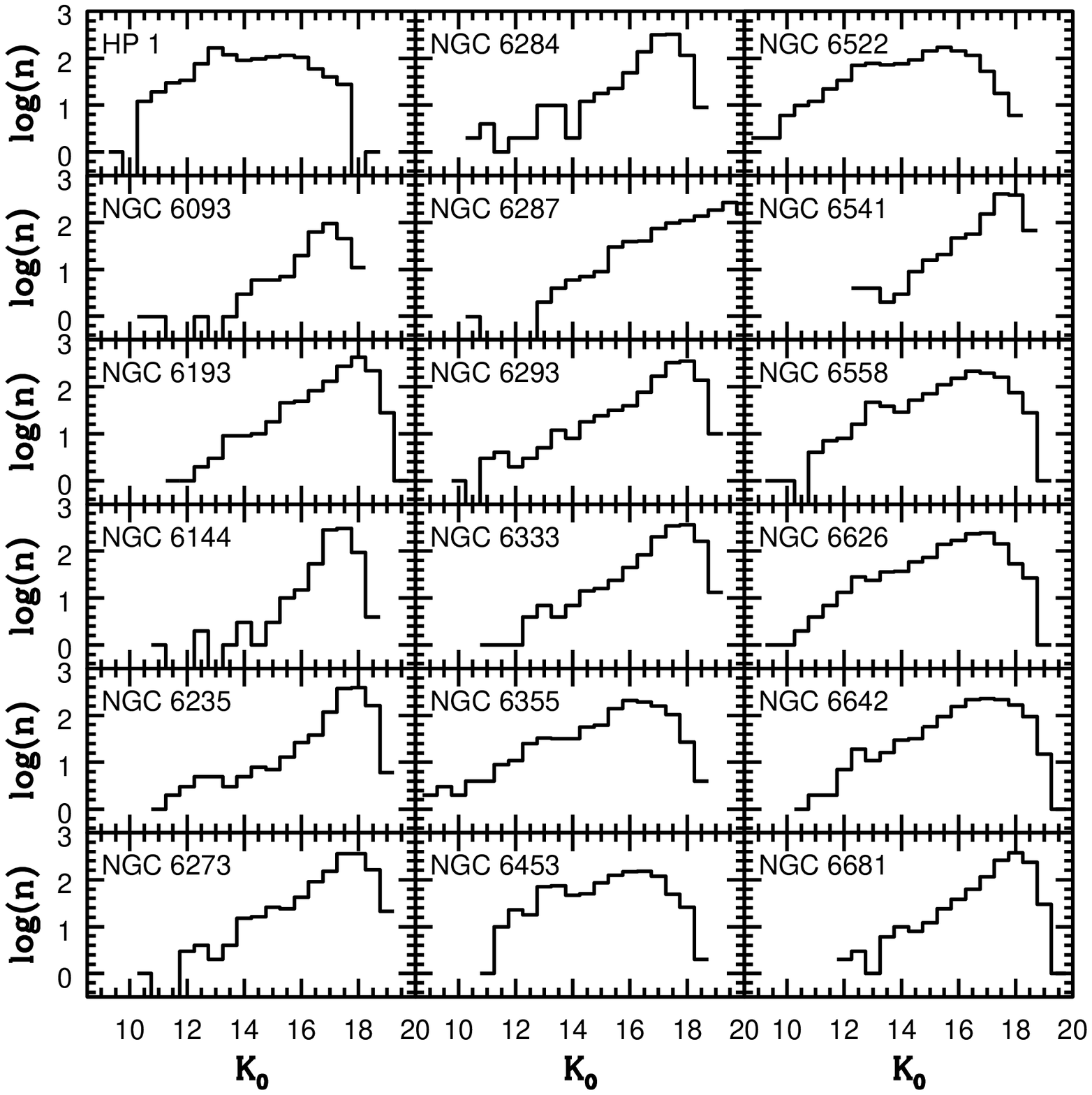]
{The reddening-corrected Field 3 LFs. $n$ is the number of stars per 0.5 mag 
interval. Bulge fields were not observed for Djorgovski 1 and NGC 6717.}

\figcaption
[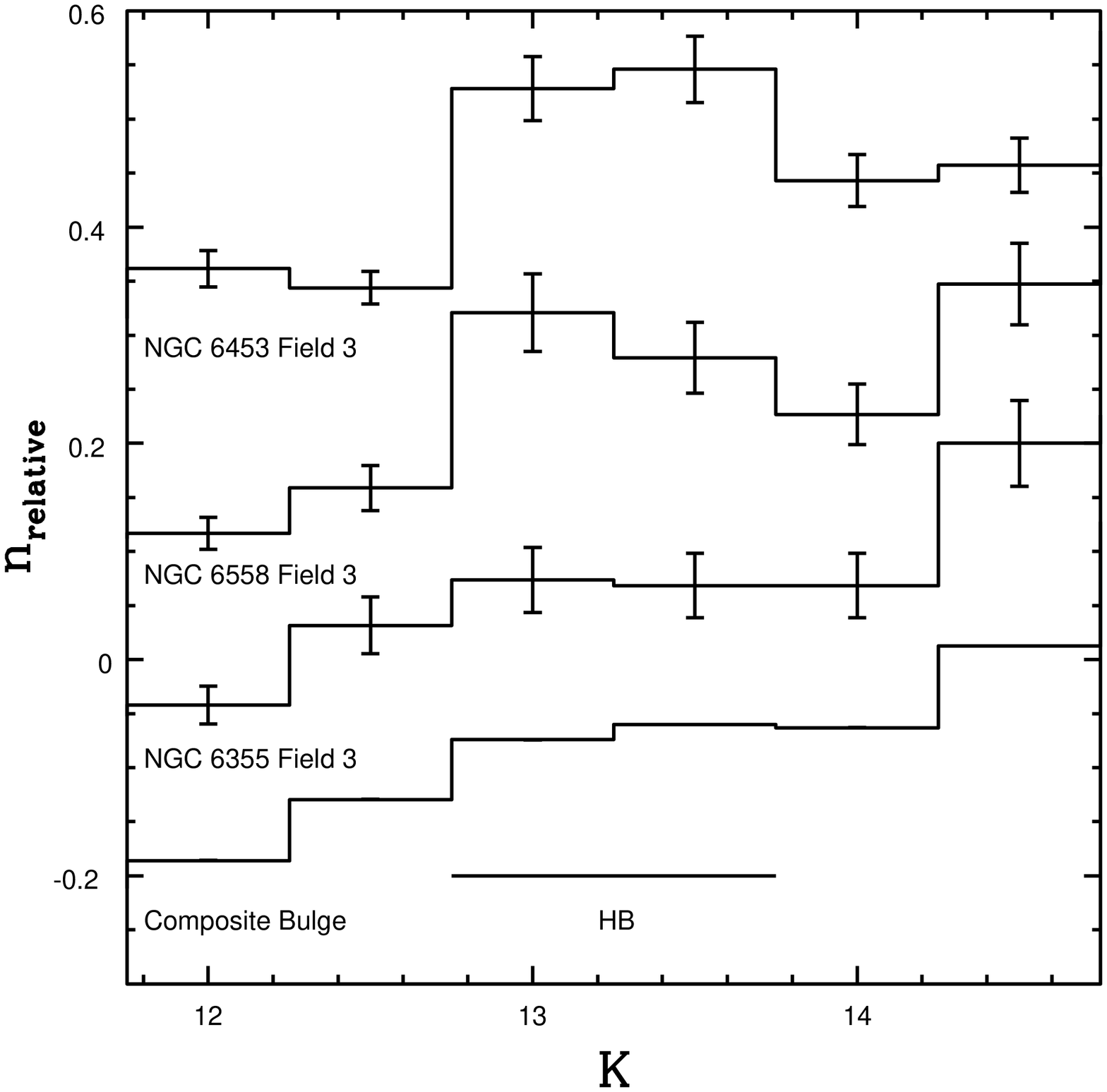]
{The $K$ LFs of NGC 6453 Field 3, NGC 6558 Field 3, NGC 6355 Field 3, 
and the composite bulge dataset in a brightness interval 
centered on the bulge HB. The error bars show the uncertainties 
in each bin based on counting statistics. The LFs have been normalized and 
shifted along the vertical axis for the purposes of this comparison. There is 
a conspicuous excess of stars in the NGC 6453 Field 3 and NGC 6558 Field 3 
LFs between $K = 12.75$ and $K = 13.75$ when compared with the NGC 6355 Field 
3 and composite bulge LFs.} 
\end{document}